\documentclass[twocolumn,journal]{IEEEtran}
\usepackage[latin9]{inputenc}
\usepackage{array}
\usepackage{amsmath}
\usepackage{amsthm}
\usepackage{amssymb}
\usepackage{graphicx}

\makeatletter

\providecommand{\tabularnewline}{\\}

\theoremstyle{plain}
\newtheorem{thm}{\protect\theoremname}
\theoremstyle{plain}
\newtheorem{lem}[thm]{\protect\lemmaname}
\theoremstyle{plain}
\newtheorem{prop}[thm]{\protect\propositionname}
\theoremstyle{plain}
\newtheorem{cor}[thm]{\protect\corollaryname}

\usepackage[caption=false,font=footnotesize]{subfig}
\allowdisplaybreaks

\usepackage{amsthm}

\newtheorem*{theorem*}{Assumption}

\@ifundefined{showcaptionsetup}{}{%
 \PassOptionsToPackage{caption=false}{subfig}}
\usepackage{subfig}
\makeatother

\providecommand{\corollaryname}{Corollary}
\providecommand{\lemmaname}{Lemma}
\providecommand{\propositionname}{Proposition}
\providecommand{\theoremname}{Theorem}

\begin{document}

\title{Cross-Mode Interference Characterization in Cellular Networks with
Voronoi Guard Regions}

\author{Stelios Stefanatos, Antonis G. Gotsis, and Angeliki Alexiou\thanks{S. Stefanatos was with the Department of Digital Systems, University
of Piraeus, Piraeus, Greece. He is now with the Department of Mathematics
and Computer Science, Freie Universit{\"a}t Berlin, Berlin, Germany
(email: stelios.stefanatos@fu-berlin.de). A. G. Gotsis and A. Alexiou
are with the Department of Digital Systems, University of Piraeus,
Piraeus, Greece (email: \{agotsis, alexiou\}@unipi.gr).}}
\maketitle
\begin{abstract}
Advances in cellular networks such as device-to-device communications
and full-duplex radios, as well as the inherent elimination of intra-cell
interference achieved by network-controlled multiple access schemes,
motivates the investigation of the cross-mode interference properties
under a guard region corresponding to the Voronoi cell of an access
point (AP). By modeling the positions of interfering APs and user
equipments (UEs) as Poisson distributed, analytical expressions for
the statistics of the cross-mode interference generated by either
APs or UEs are obtained based on appropriately defined density functions.
The considered system model and analysis are general enough to capture
many operational scenarios of practical interest, including conventional
downlink/uplink transmissions with nearest AP association, as well
as transmissions where not both communicating nodes lie within the
same cell. Analysis provides insights on the level of protection offered
by a Voronoi guard region and its dependence on type of interference
and receiver position. Numerical examples demonstrate the validity/accuracy
of the analysis in obtaining the system coverage probability for operational
scenarios of practical interest. 
\end{abstract}

\begin{IEEEkeywords}
Stochastic geometry, cellular networks, guard region, D2D communications,
full-duplex radios, interference.
\end{IEEEkeywords}

\section{Introduction}

Characterization of the interference experienced by the receivers
of a wireless network is of critical importance for system analysis
and design \cite{Haenggi=000026Ganti}. This is especially the case
for the future cellular network, whose envisioned fundamental changes
in its architecture, technology, and operation will have significant
impact on the interference footprint \cite{Boccardi 5 disruptive}.
Interference characterization under these new system features is of
the utmost importance in order to understand their potential merits
as well as their ability to co-exist.

Towards increasing the spatial frequency reuse, two of the most prominent
techniques/features considered for the future cellular network are
device-to-device (D2D) communications \cite{D2D general} and full-duplex
(FD) radios \cite{FD JSAC tutorial}. Although promising, application
of these techniques introduces additional, \emph{cross-mode} interference.
For example, an uplink transmission is no longer affected only by
interfering uplink transmissions but also by interfering (inband)
D2D and/or downlink transmissions. Although it is reasonable to expect
that the current practice of eliminating intra-cell interference by
employing coordinated transmissions per cell will also hold in the
future \cite{D2D mag network controlled}, the continuously increasing
density of APs and user equipments (UEs) suggest that inter-cell interference
will be the major limiting factor of D2D- and/or FD-enabled cellular
networks, rendering its statistical characterization critical.

\subsection{Previous Work}

Stochastic geometry is by now a well accepted framework for analytically
modeling interference in large-scale wireless networks \cite{JSAC stoc geom}.
Under this framework, most of the numerous works on D2D-enabled cellular
networks (without FD radios) consider the interfering D2D nodes as
uniformly distributed on the plane, i.e., there is no spatial coordination
of D2D transmissions (see, e.g., \cite{D2D Ye,D2D Lin,D2D Lee,Stefanatos}).
Building on the approach of \cite{Andrews guard zone}, various works
consider the benefits of employing \emph{spatially coordinated} D2D
transmissions where, for each D2D link in the system, a circular guard
region (zone) is established, centered at either the receiver or transmitter,
within which no interfering transmissions are performed \cite{George D2D,Kountouris D2D}.
However, when the D2D links are network controlled \cite{D2D mag network controlled},
a more natural and easier to establish guard region is the (Voronoi)
cell of a coordinating AP. Under a non-regular AP deployment \cite{Primer on PPP},
this approach results in a random polygon guard region, which makes
the interference characterization a much more challenging task. 

Interference characterization for this type of guard region has only
been partially investigated in \cite{ElSawy uplink,UL Comm Lett,Andrews joint UL-DL}
and only for the case of conventional uplink transmissions with nearest
AP association and one active UE per cell (with no cross-mode interference).
As the positions of the interfering UEs are distributed as a Voronoi
perturbed lattice process (VPLP) in this case \cite{PVLP,Haenggi PVLP},
which is analytically intractable, an approximation based on a Poisson
point process (PPP) model with a \emph{heuristically} proposed \emph{equivalent}
\emph{density} is employed. This approach of approximating a complicated
system model by a simpler one with appropriate parameters (in this
case, by a PPP of a given density) was also used in \cite{DiRenzo}
for the characterization of downlink systems (with no cross-mode interference
as well). Reference \cite{Haenggi PVLP} provides a rigorous characterization
of the equivalent density of the UE-generated interference both from
the ``viewpoint'' of an AP as well as its associated UE, with the
latter case of interest in case of cross-mode interference. The analysis
reveals significant differences in the equivalent densities corresponding
to these two cases suggesting that the equivalent density is strongly
dependent on the considered receiver position. Interference characterization
for the case of arbitrary receiver position that may potentially lie
outside the Voronoi guard region as, e.g., in the case of cross-cell
D2D links \cite{cross-cell D2D}, has not been investigated in the
literature, let alone under cross-mode interference conditions.

Investigation of interference statistics under a cell guard region
is also missing in the (much smaller) literature on FD-enabled cellular
networks, which typically assumes no spatial coordination for the
UE transmissions (see e.g., \cite{Tong=000026Haengi FD,Psomas,ElSawy FD with D2D}).

\subsection{Contributions}

This paper considers a stochastic geometry framework for modeling
the \emph{cross-mode} interference power experienced at an \emph{arbitrary
receiver position} due to transmissions by APs or UEs and under the
spatial protection of a Voronoi guard region. Modeling the positions
of interfererers (APs or UEs) as a Poisson point process, the statistical
characterization of the cross-mode interference power is pursued via
computation of its Laplace transform. The main contributions of the
paper are the following.
\begin{itemize}
\item Consideration of a general system model, which allows for a unified
analysis analysis of cross-mode interference statistics. By an appropriate
choice of the system model parameters, the interference characterization
is applicable to multiple operational scenarios, including conventional
downlink/uplink communications with nearest AP association and spatially
coordinated (cross-cell) D2D links where the transmitter-receiver
pair does not necessarily lie in a single cell.
\item Exact statistical characterization of AP-generated cross-mode interference,
applicable, e.g., in the case where interference experienced at an
AP is due to transmissions by other FD-enabled APs. An equivalent
interferer density is given in a simple closed form, allowing for
an intuitive understanding of the interfernece properties and its
dependence on the position of the receiver relative to the position
of the AP establishing the Voronoi guard region. 
\item Determination of a lower bound for the Laplace transform of UE-generated
cross-mode interference power, applicable, e.g., in the case where
a UE experiences interference due to other FD-enabled or D2D-operating
UEs. The properties of the corresponding equivalent density of interferers
is studied in detail, providing insights for various operational scenarios,
including a rigorous justification of why the heuristic approaches
previously proposed in \cite{UL Comm Lett,Andrews joint UL-DL} for
the analysis of the standard uplink communication scenario (with no
cross-mode interfernece) are accurate.
\end{itemize}
Simulated examples indicate the accuracy of the analytical results
also for cases where the positions of interferering UEs are VPLP distributed,
suggesting their use as a basis for determination of performance as
well as design of optimal resource allocation algorithms for future,
D2D- and/or FD-enabled cellular networks.

\subsection{Notation}

The origin of the two-dimensional plane $\mathbb{R}^{2}$ will be
denoted as $o$. The Euclidean norm of $x\in\mathbb{R}^{2}$ is denoted
as $|x|$ with operator $|\cdot|$ also used to denote the absolute
value of a scalar or the Lebesgue measure (area) of a bounded subset
of $\mathbb{R}^{2}$. The polar form representation of $x\in\mathbb{R}^{2}$
will be denoted as $(|x|,\angle x$) or $|x|\angle x$, where $\angle x$
is the principal branch of the angular coordinate of $x$ taking values
in $[-\pi,\pi)$. The open ball in $\mathbb{R}^{2}$, centered at
$x\in\mathbb{R}^{2}$ and of radius $R>0$, is denoted as $\mathcal{B}(x,R)\triangleq\{y\in\mathbb{R}^{2}:|y-x|<R\}$,
whereas its boundary is denoted as $\mathcal{C}(x,R)\triangleq\{y\in\mathbb{R}^{2}:|y-x|=R\}$.
$\mathbb{I}(\cdot)$ is the indicator ($0-1$) operator, $\mathbb{P}(\cdot)$
is the probability measure, and $\mathbb{E}(\cdot)$ is the expectation
operator. Functions $\arccos(\cdot):[-1,1]\rightarrow[0,\pi]$ and
$\arcsin(\cdot):[-1,1]\rightarrow[-\pi/2,\pi/2]$ are the principal
branches of the inverse cosine and sine, respectively. The Laplace
transform of a random variable $z$ equals $\mathcal{L}_{z}(s)\triangleq\mathbb{E}(e^{-sz}),$
for all $s\in\mathbb{R}$ for which the expectation exists.

\section{System Model}

A large-scale model of a cellular network with APs and UEs positioned
over $\mathbb{R}^{2}$ is considered. The positions of APs are modeled
as a realization of a homogeneous PPP (HPPP) $\Phi_{a}\subset\mathbb{R}^{2}$
of density $\lambda_{a}>0$. All the communication scenarios considered
in this paper involve three nodes, namely,
\begin{itemize}
\item a, so called, typical node, which, without loss of generality (w.l.o.g.),
will be assumed that its position coincides with the origin $o$.
The typical node is an AP if $o\in\Phi_{a}$ or a UE, otherwise
\item the closest AP to the typical node, located at $x^{*}\triangleq\arg\min_{x\in\Phi_{a}}|x|$
($x^{*}=o$ in case $o\in\Phi_{a}$, i.e., the typical node is an
AP);
\item a receiver located at an arbitrary position $x_{R}\in\mathbb{R}^{2}$.
The receiver is an AP if $x_{R}\in\Phi_{a}$ or a UE, otherwise. The
case $x_{R}=o$ is also allowed, meaning that the typical node is
also the receiver.
\end{itemize}
Let $\mathcal{V}^{*}$ denote the Voronoi cell of the AP at $x^{*}$,
generated by the Poisson-Voronoi tessellation of the plane from $\Phi_{a}$,
i.e., 
\[
\mathcal{V}^{*}\triangleq\left\{ y\in\mathbb{R}^{2}:|y-x^{*}|<|y-x|\text{, for all }x\in\Phi_{a}\setminus\{x^{*}\}\right\} .
\]
The goal of this paper is to characterize the interference power experienced
at $x_{R}$ due to transmissions from nodes (APs or UEs) that lie
outside $\mathcal{V}^{*}$, i.e., with $\mathcal{V}^{*}$ effectively
forming a spatial guard region within which no interference is generated.

\begin{figure}
\noindent \centering{}\includegraphics[width=0.6\columnwidth]{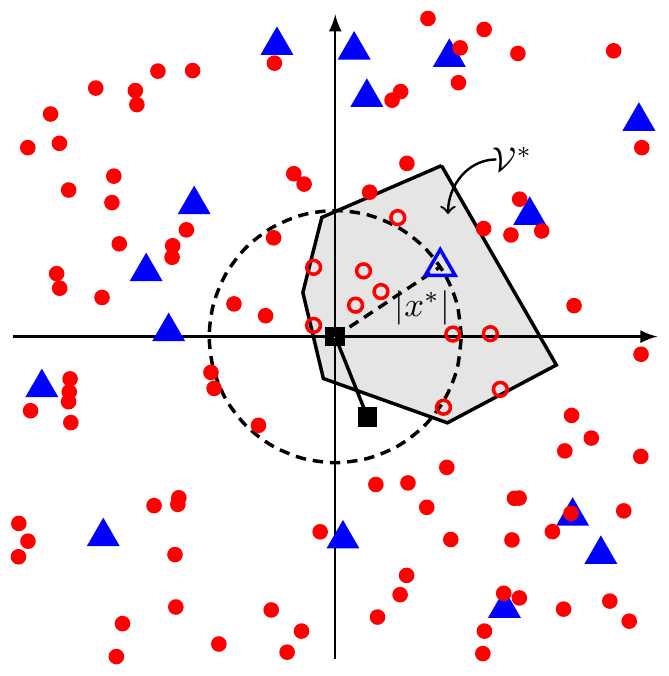}\caption{\label{fig:system model}Random realization of the system model. APs
and UEs are shown as triangle and circle markers, respectively. The
typical node in this example is also a UE shown as a square. The shaded
polygon area indicates the Voronoi guard region $\mathcal{V}^{*}$
within which no node (indicated by open marker) transmits. All nodes
outside $\mathcal{V}^{*}$ (filled markers) generate interference
that is experienced by the receiver of interest whose position ($x_{R})$
is indicated by an open diamond marker. This scenario may correspond
to a (cross-cell) D2D link between typical node and receiver or to
the receiver acting as relay aiding the donwlink or uplink communication
of the typical node with its nearest AP.}
\end{figure}

Let $\Phi_{u}\subset\mathbb{R}^{2}$ denote the point process representing
the positions of the UEs in the system other than the typical node
and receiver, and $I_{x_{R},a}$ and $I_{x_{R},u}$ denote the interference
power experienced at $x_{R}$ due to transmissions by all APs and
UEs in the system, respectively. Under the Voronoi guard region scheme
described above, the standard interference power model is adopted
in this paper, namely \cite{Haenggi=000026Ganti},
\begin{equation}
I_{x_{R},k}\triangleq\sum_{x\in\Phi_{k}\setminus\mathcal{V}^{*}}P_{k}g_{x}|x-x_{R}|^{-\alpha_{k}},k\in\{a,u\},\label{eq:I definition-2}
\end{equation}

\noindent where $g_{x}\geq0$ is the channel gain of a transmission
generated by a node at $x\in\mathbb{R}^{2}$, assumed to be an independent
of $x$, exponentially distributed random variable of mean $1$ (Rayleigh
fading), $\alpha_{k}>2$ is the path loss exponent and $P_{k}>0$
is the transmit power, which, for simplicity, is assumed fixed and
same for all nodes of the same type. Figure \ref{fig:system model}
shows an example realization of $\mathcal{V}^{*}$ and the positions
of the interferers for the case $x_{R}\neq x^{*}\neq o$. Note that
$\mathcal{V}^{*}$ is a \emph{random} guard region as it depends on
the positions of the APs, therefore, there is no guarantee that $x_{R}$
lies within $\mathcal{V}^{*}$, i.e., it holds $\mathbb{P}(x_{R}\in\mathcal{V}^{*})<1$,
unless $x_{R}$ is a point on the line segment joining the origin
with $x^{*}$ (inclusive).

The above model is general enough so that, by appropriately choosing
$x_{R}$ and/or $x^{*}$, $I_{x_{R},a}$ and $I_{x_{R},u}$ correspond
to many practical instances of (cross-mode) interference experienced
in D2D/FD-enabled as well as conventional cellular networks. For example,
for the case where $x^{*}\neq x_{R}\neq o$, Eq. (\ref{eq:I definition-2})
may correspond to the AP- and UE-generated interference power experienced
by the receiver of a, potentially cross-cell, D2D link between the
typical nodes at $o$ and $x_{R}$. Various other scenarios of practical
interest are captured by the model, some of which are described in
Table \ref{tab:Special-Cases-of}. Note that from the scenarios identified
in Table \ref{tab:Special-Cases-of}, only the standard downlink and
uplink scenarios have been considered previously in the literature
\cite{ElSawy uplink,UL Comm Lett,Andrews joint UL-DL,Andrews tractable},
however, \emph{without consideration of cross-mode interference},
i.e., UE-generated and AP-generated interference for downlink and
uplink transmissions, respectively. Reference \cite{Haenggi PVLP}
considers the UE-generated cross-mode interference experienced at
the typical node in downlink mode, however, the results provided cannot
not be straightforwardly generalized to the general case.

\begin{table}[t]
\noindent \centering{}\caption{\label{tab:Special-Cases-of}Special Cases of System Model}
\begin{tabular}{|>{\centering}m{0.14\columnwidth}|>{\centering}m{0.75\columnwidth}|}
\hline 
condition & scenario\tabularnewline
\hline 
\hline 
$x_{R}\neq x^{*}$,

$x^{*}\neq o$ & General case. Corresponds to (a) a D2D link when the typical node
sends data to the receiver or (b) to a relay-assisted cellular communication
where the receiver acts a relay for the link between the typical node
and it nearest AP (either downlink or uplink). The receiver is not
guaranteed to lie within $\mathcal{V}^{*}$.\tabularnewline
\hline 
$x_{R}=o$, $x^{*}\neq o$ & Typical node is a UE in receive mode and lies within $\mathcal{V}^{*}$.
Represents a standard downlink when the nearest AP is transmitting
data to the typical node/receiver. \tabularnewline
\hline 
$x_{R}=x^{*}$, $x^{*}\neq o$ & Nearest AP to the typical node is in receive mode with the typical
node lying in $\mathcal{V}^{*}$. Represents a standard uplink when
the AP receives data from the typical node.\tabularnewline
\hline 
$x_{R}\neq o$, $x^{*}=o$ & Typical node is an AP. When the typical node/AP is the one transmitting
to the receiver, a non-standard downlink is established as $x_{R}$
is not necessarily lying within $\mathcal{V}^{*}$.\tabularnewline
\hline 
$x_{R}=x^{*}$,

$x^{*}=o$ & Typical node is an AP in receive mode; the position of the corresponding
transmitter is unspecified and may as well lie outside $\mathcal{V}^{*}$,
thus modeling a non-standard uplink.\tabularnewline
\hline 
\end{tabular}
\end{table}
For characterization of the performance of the communication link
as well as obtaining insights on the effectiveness of the guard region
$\mathcal{V}^{*}$ for reducing interference, it is of interest to
describe the statistical properties of the random variables $I_{x_{R},a}$
and $I_{x_{R},u}$ . This is the topic of the following section, where
the marginal statistics of $I_{x_{R},a}$ and $I_{x_{R},u}$\emph{
conditioned on $x^{*}$ and} \emph{treating $x_{R}$ as a given, but
otherwise free, parameter}, are investigated in detail. Characterization
of the joint distribution of $I_{x_{R},a}$ and $I_{x_{R},u}$ is
left for future work. 

Note that the respective unconditional interference statistics can
be simply obtained by averaging over the distribution of $x^{*}$,
which corresponds to a uniformly distributed $\angle x^{*}$ and a
Rayleigh distributed $|x^{*}|$ of mean $1/(2\sqrt{\lambda_{a}})$
\cite{Baccelli}. However, results conditioned on $x^{*}$ are also
of importance on their own as they can serve as the mathematical basis
for (optimal) resource allocation algorithms given network topology
information, e.g., decide on whether the typical node employs D2D
or cellular mode given knowledge of $x^{*}$.

For tractability purposes, the following assumption on the statistics
of $\Phi_{u}$ will be considered throughout the analysis.

\begin{theorem*} 
The positions of the (potentially) interfering UEs, $\Phi_u$, is a realization of an HPPP of density  $\lambda_u>0$, that is independent of the positions of the APs, typical node, and  receiver.
\end{theorem*}

\emph{Remark}: In general, this assumption is not exact since resource
allocation and scheduling decisions over the entire network affect
the distribution of interfering UE positions. For example, in the
conventional uplink scenario with at least one UE per cell, the transmitting
UE positions correspond to a VPLP process of density equal to $\lambda_{u}=\lambda_{a}$
\cite{Andrews joint UL-DL,Haenggi PVLP}. However, as also shown in
\cite{Andrews joint UL-DL,Haenggi PVLP} for the standard uplink scenario,
the HPPP assumption of the UE point process allows for a tractable,
yet accurate approximation of the actual performance, which, as will
be demonstrated in Sec. V, is also the case for other operational
scenarios as well. For generality purposes, an arbitrary value $\lambda_{u}\neq\lambda_{a}$
is also allowed in the analysis, which can actually the case, e.g.,
in ultra dense networks with at most one UE allowed to transmit per
cell, resulting in $\lambda_{u}<\lambda_{a}$ \cite{Stefanatos UDN},
and also serves as an approximate model for the case when some arbitrary/undefined
coordination scheme is employed by other cells in the system (if at
all), that may as well result in $\lambda_{u}>\lambda_{a}$. 

\section{Interference Characterization}

Towards obtaining a tractable characterization of the distribution
of the (AP or UE) interference power, or, equivalently, its Laplace
transform, the following standard result in PPP theory is first recalled
\cite{Baccelli}.
\begin{lem}
\label{lem: Laplace Poisson theory}Let $I\triangleq\sum_{x\in\tilde{\Phi}}Ph_{x}|x-z|^{-\alpha},z\in\mathbb{R}^{2}$,
$P>0$, with $\tilde{\Phi}$ an inhomogeneous PPP of density $\lambda:\mathbb{R}^{2}\rightarrow[0,\infty)$,
and $\{h_{x}\}_{x\in\Phi}$ i.i.d. exponential random variables of
mean $1$. The Laplace transform of $I$ equals
\begin{align}
\mathcal{L}_{I}(s) & =\exp\left\{ -\int_{\mathbb{R}^{2}}\lambda(x)\gamma(sP|x-z|^{-\alpha})dx\right\} \label{eq:2D integral Laplace transform}\\
 & =\exp\left\{ -2\pi\int_{0}^{\infty}\lambda_{z}(r)r\gamma(sPr^{-\alpha})dr\right\} ,\label{eq:radial Laplace transform}
\end{align}

\noindent where $\gamma(t)\triangleq1-\frac{1}{1+t}$, and the second
equality holds only in the case of a circularly-symmetric density
function w.r.t. point $z$, i.e., $\lambda(z+x)=\lambda_{z}(|x|)$,
for an appropriately defined function $\lambda_{z}:[0,\infty)\rightarrow[0,\infty)$.
\end{lem}
The above result provides a complete statistical characterization
in integral form of the interference experienced at a position $z\in\mathbb{R}^{2}$
due to Poisson distributed interferers. Of particular interest is
the case of a circularly-symmetric density, which only requires a
single integration over the radial coordinate and has previously led
to tractable analysis for various wireless network models of interest
such as mobile \emph{ad hoc} \cite{Haenggi=000026Ganti} and downlink
cellular \cite{Andrews tractable}. In the following, it will be shown
that even though the interference model of Sec. II suggests a non
circularly-symmetric interference density due to the random shape
of the guard region, the Laplace transform formula of (\ref{eq:radial Laplace transform})
holds exact for $I_{x_{R},a}$ and is a (tight) lower bound for $I_{x_{R},u}$
with appropriately defined circularly-symmetric equivalent density
functions.

\subsection{AP-Generated Interference}

Note that considering the Voronoi cell of a \emph{random} AP at $x\in\Phi_{a}$
acting as a guard region has no effect on the distribution of the
interfering APs. This is because the Voronoi cell of any AP is a (deterministic)
function of $\Phi_{a}$, therefore, it does not impose any constraints
on the realization of $\Phi_{a}$, apart from implying that the AP
at $x$ does not produce interference. By well known properties of
HPPPs \cite{Baccelli}, conditioning on the position $x$ of the non-interfering
AP, has no effect on the distribution of the interfering APs. That
is, the interfering AP positions are still distributed as an HPPP
of density $\lambda_{a}$ as in the case without imposing any guard
region whatsoever.

\begin{figure}
\noindent \centering{}\includegraphics[width=1\columnwidth]{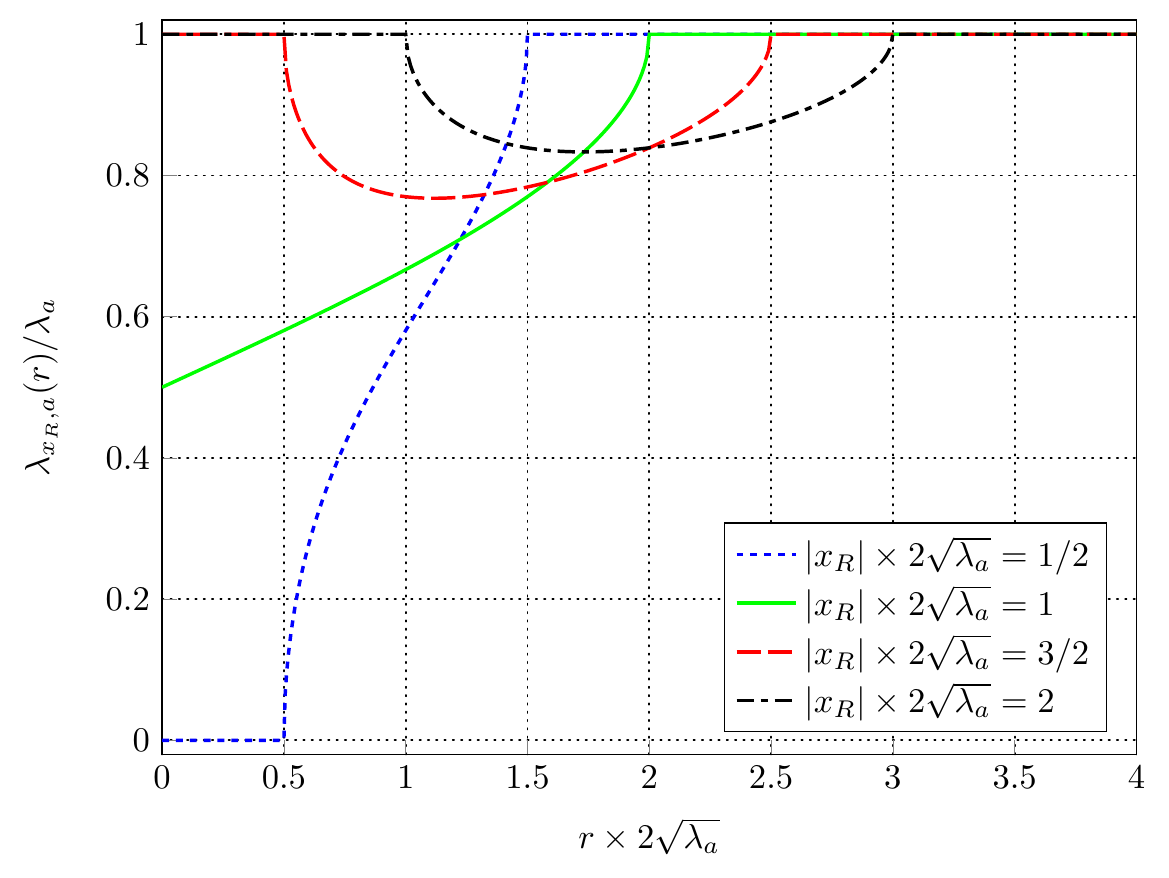}\caption{\label{fig:AP_to_RL_density_vs_rho_RL}Equivalent radial density of
interfering APs experienced at various $x_{R}\in\mathbb{R}^{2}$ ,
conditioned on $|x^{*}|=1/(2\sqrt{\lambda_{a}})$.}
\end{figure}

However, when it is the Voronoi cell of the nearest AP to the origin
that is considered as a guard region, an \emph{implicit} guard region
w.r.t. AP-generated interference is formed. Indeed, the interfering
APs, given the AP at $x^{*}$ is not interfering, are effectively
distributed as an inhomogeneous PPP with density \cite{Andrews tractable}
\begin{equation}
\tilde{\lambda}_{a}(x)=\lambda_{a}\mathbb{I}(x\notin\mathcal{B}(o,|x^{*}|)),x\in\mathbb{R}^{2},\label{eq:andrews_density}
\end{equation}
 i.e., an implicit circular guard zone is introduced around the typical
node (see Fig. \ref{fig:system model}), since, if this was not the
case, another AP could be positioned at a distance smaller than $|x^{*}|$
from the origin, which is impossible by assumption. This observation
may be used to obtain the Laplace transform of the AP-generated interference
experienced at $x_{R}$ directly from the formula of (\ref{eq:2D integral Laplace transform}).
However, the following result shows that the two-dimensional integration
can be avoided as the formula of (\ref{eq:radial Laplace transform})
is also valid in this case with an appropriately defined equivalent
density function.
\begin{prop}
\label{prop:type-I Laplace}The Laplace transform, $\mathcal{L}_{I_{x_{R},a}}(s\mid x^{*})\triangleq\mathbb{E}(e^{-sI_{x_{R},a}}\mid x^{*})$,
of the AP-generated interference power $I_{x_{R},a}$, conditioned
on $x^{*}$, equals the right-hand side of (\ref{eq:radial Laplace transform})
with $P=P_{a}$, $\alpha=\alpha_{a}$ and $\lambda_{z}(r)=\lambda_{x_{R},a}(r),r\geq0$,
where
\begin{equation}
\lambda_{x_{R},a}(r)\triangleq\begin{cases}
\lambda_{a} & ,r>|x^{*}|+|x_{R}|\\
\begin{cases}
0 & ,|x^{*}|>|x_{R}|,\\
\lambda_{a} & ,|x^{*}|<|x_{R}|,
\end{cases} & ,r\leq\left||x^{*}|-|x_{R}|\right|\\
\lambda_{a}/2 & ,|x^{*}|=|x_{R}|\\
\lambda_{a}\left(1-\frac{1}{\pi}\arccos\left(d\right)\right) & ,\text{\emph{otherwise}},
\end{cases}\label{eq:AP-interferer_density}
\end{equation}

\noindent with $d\triangleq\frac{r^{2}-(|x^{*}|^{2}-|x_{R}|^{2})}{2r|x_{R}|}$,
for $x_{R}\neq o$, and
\begin{equation}
\lambda_{x_{R},a}(r)=\lambda_{a}\mathbb{I}(r\geq|x^{*}|),\label{eq:DL AP density}
\end{equation}

\noindent for $x_{R}=o$.
\end{prop}
\begin{IEEEproof}
See Appendix A.
\end{IEEEproof}
The following remarks can be made:
\begin{enumerate}
\item The interference power experienced at an arbitrary position $x_{R}\neq o$
under the considered guard region scheme is equal in distribution
to the interference power experienced at the origin without any guard
region and with interferers distributed as an inhomogeneous PPP of
an equivalent, $x_{R}$-depednent density given by (\ref{eq:AP-interferer_density}).
\item Even though derivation of the statistics of $I_{x_{R},a}$ was conditioned
on $x_{R}$ and $x^{*}$, the resulting equivalent inteferer density
depends only on their norms $|x_{R}|$ and $|x^{*}|$. Although the
indepedence from $\angle x^{*}$ might have been expected due to the
isotropic property of $\Phi_{a}$ \cite{Haenggi=000026Ganti}, there
is no obvious reason why one would expect independence also from $\angle x_{R}$.
\item $\lambda_{x_{R},a}(r)$ is a decreasing function of $|x^{*}|$, corresponding
to a (statistically) smaller AP interference power due to an increased
guard zone area.
\item For $x_{R}=o$, corresponding to a standard downlink with interference
generated from other donwlink transmissions, Prop. \ref{prop:type-I Laplace}
coincides with the analysis of \cite{Andrews tractable}, as expected.
\item The Laplace transform of the interference in the case of $x_{R}\neq x^{*}\neq o$
was examined previously in \cite{Kountouris 2}. However, the corresponding
formulas appear as two-dimensional integrals that offer limited insights
compared to the simpler and more intuitive equivalent density formulation
given in Prop. \ref{prop:type-I Laplace}.
\item The case $x_{R}=x^{*}\neq o$ corresponds to the nearest-neighbor
transmission scenario considered in \cite{Haenggi local delay} where
the validity of $\lambda_{x_{R},a}(r)$ in (\ref{eq:AP-interferer_density})
as an equivalent density function for computation of the Laplace transform
of the interference power was not observed.
\item In \cite{Dhillon hole}, the Laplace transform of the interference
power experienced at $x_{R}\in\mathbb{R}^{2}$ due to a Poisson hole
process, i.e., with interferers distributed as an HPPP over $\mathbb{R}^{2}$
except in the area covered by randomly positioned disks (holes), was
considered. The holes were assumed to not include $x_{R}$ and a lower
bound for the Laplace transform was obtained by considering only a
single hole \cite[Lemma 5]{Dhillon hole}, which coincides with the
result of Prop. \ref{prop:type-I Laplace}. This is not surprising
as the positions of the APs, conditioned on $x^{*}$, are essentially
distributed a single-hole Poisson process. Note that Prop. \ref{prop:type-I Laplace}
generalizes \cite[Lemma 5]{Dhillon hole} by allowing $x_{R}$ to
be covered by the hole and considers a different proof methodology.
\end{enumerate}
Figure \ref{fig:AP_to_RL_density_vs_rho_RL} shows the normalized
density function $\lambda_{x_{R},a}(r)/\lambda_{a}$ for various values
of $|x_{R}|$ and assuming that $|x^{*}|=1/(2\sqrt{\lambda_{a}})$,
the expected distance from the nearest AP. It can be seen that the
presence of the implicit circular guard region results in reducing
the equivalent interferer density in certain intervals of the radial
coordinate $r$, depending on the value of $|x_{R}|$. In particular,
when $|x_{R}|<|x^{*}|$, it is guaranteed that no APs exist within
a radius $|x^{*}|-|x_{R}|>0$ from $x_{R}$. In contrast, when $|x_{R}|>|x^{*}|$
there is no protection from APs in the close vicinity of $x_{R}$. 

\begin{figure}
\noindent \centering{}\includegraphics[width=1\columnwidth]{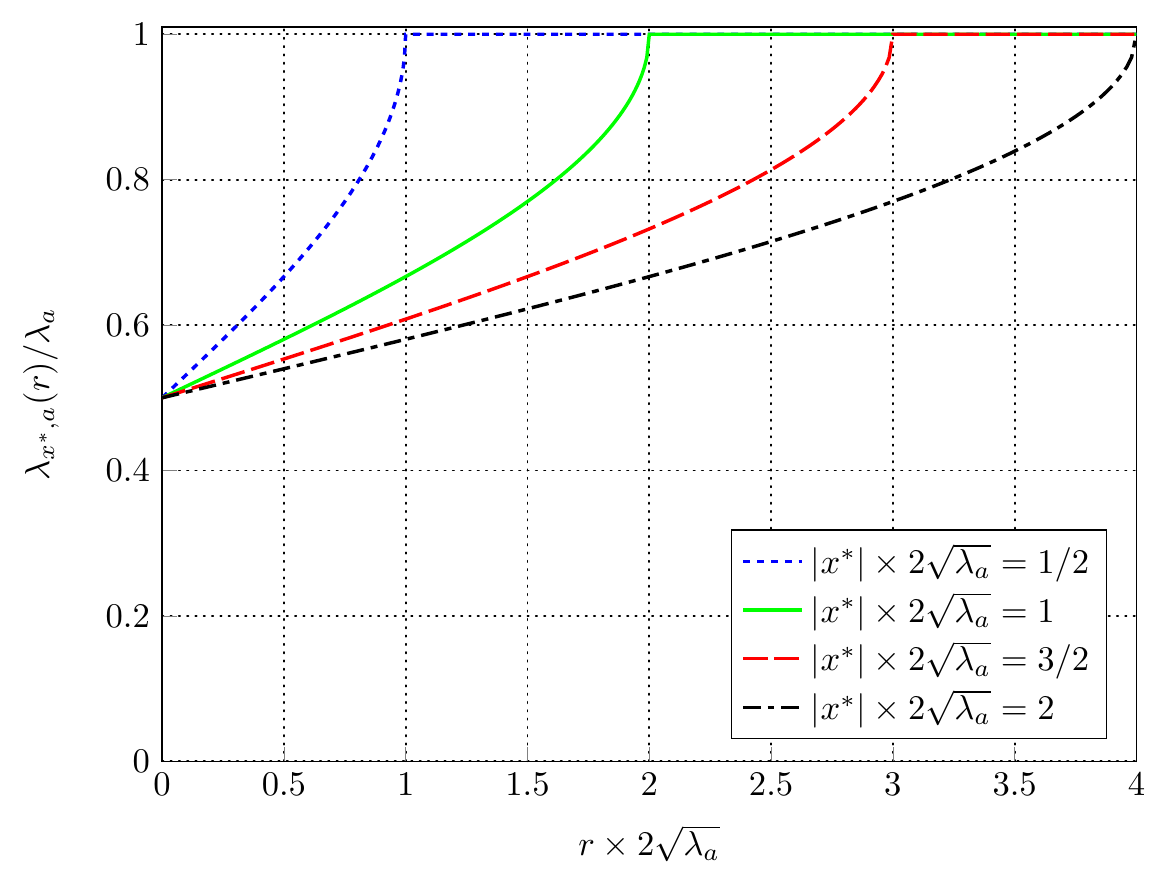}\caption{\label{fig:AP_to_AP_density_vs_rho_AP}Equivalent radial density of
interfering APs experienced at $x^{*}$.}
\end{figure}

The case $|x_{R}|=|x^{*}|$ is particularly interesting since it corresponds
to the case when the receiver is the AP at $x^{*}$, experiencing
interference from other AP, e.g., when operating in FD. For $x^{*}\neq o$,
corresponding to an uplink transmission by the typical node to its
nearest AP, it can be easily shown that it holds
\begin{equation}
\lambda_{x^{*},a}(r)=\lambda_{a}\left(\frac{1}{2}+\frac{r}{2\pi|x^{*}|}\right)+\mathcal{O}(r^{3}),r\rightarrow0,\label{eq:asymptitic_AP_to_AP_density}
\end{equation}

\noindent i.e., the guard region results in the serving AP experiencing
about half of the total interfering APs density in its close vicinity,
which is intuitive as for asymptotically small distances from $x^{*}$
the boundary of the circular guard region in (\ref{eq:andrews_density})
can be locally approximated as a line that divides the plane in two
halves, one with interferer density $\lambda_{a}$ and one with interferer
density $0$. Figure \ref{fig:AP_to_AP_density_vs_rho_AP} shows the
normalized $\lambda_{x^{*},a}(r)$ for various values of $|x^{*}|$,
where its linear asymptotic behavior as well as the advantage of a
larger $|x^{*}|$ are clearly visible.

\subsection{UE-Generated Interference}

The positions of interfering UEs, conditioned on the realization of
$\Phi_{a}$, are distributed as an inhomogeneous PPP of density
\begin{equation}
\tilde{\lambda}_{u}(x|\Phi_{a})=\lambda_{u}\mathbb{I}(x\notin\mathcal{V}^{*}),x\in\mathbb{R}^{2}.\label{eq:UE density general}
\end{equation}

\noindent Although the expression for $\tilde{\lambda}_{u}$ is very
similar to the that of the density $\tilde{\lambda}_{a}$ of interfering
APs given in (\ref{eq:andrews_density}), the random Voronoi guard
region $\mathcal{V}^{*}$ appearing in (\ref{eq:UE density general})
is significantly more complicated than the deterministic circular
guard zone $\mathcal{B}(o,|x^{*}|)$, which renders the analysis more
challenging. To simplify the following exposition, it will be assumed
that a rotation of the Cartesian coordinate axes is performed such
that $\angle x^{*}=0$. Note that this rotation has no effect in the
analysis due to the isotropic property of the HPPP \cite{Haenggi=000026Ganti}
and immediately renders the following results independent of the value
of $\angle x^{*}$ in the original coordinate system.

Since it is of interest to examine the UE interference statistics
conditioned only on $x^{*}$, a natural quantity to consider, that
will be also of importance in the interference statistics analysis,
is the \emph{probabilistic cell area} (PCA) function $p_{c}(x\mid x^{*}),x\in\mathbb{R}^{2}$.
This function gives the probability that a point $x\in\mathbb{R}^{2}$
lies within $\mathcal{V}^{*}$ conditioned only on $x^{*}$, i.e.,\footnote{Recall that, under the system model, $x^{*}$ is also the AP closest
to the origin.} 
\begin{align*}
p_{c}(x\mid x^{*}) & \triangleq\mathbb{\mathbb{P}}(x\in\mathcal{V}^{*}\mid x^{*}).
\end{align*}

\begin{lem}
\label{lem:PCA}For all $x\in\mathbb{R}^{2}$, the PCA function equals
\[
p_{c}(x\mid x^{*})=\begin{cases}
1 & ,|x|\leq|x^{*}|,\angle x=0,\\
e^{-\lambda_{a}|\mathcal{A}|}<1 & ,\text{\emph{otherwise}},
\end{cases}
\]

\noindent where $\mathcal{A}\triangleq\mathcal{B}(x,|x-x^{*}|)\setminus\mathcal{B}(o,|x^{*}|)$.
For all $x\neq x^{*}$, it holds
\[
|\mathcal{A}|\!=\!\begin{cases}
r_{*}^{2}(|\angle x|\!+\!\theta_{*})\!-\!|x^{*}|^{2}|\angle x|\!+\!|x||x^{*}|\sin(|\angle x|)\! & ,x^{*}\neq o,\\
\pi|x|^{2}\negmedspace & ,x^{*}=o,
\end{cases}
\]

\noindent with $r_{*}\triangleq|x-x^{*}|=\sqrt{|x|^{2}+|x^{*}|^{2}-2|x||x^{*}|\cos(\angle x)}$
and 
\[
\theta_{*}\triangleq\begin{cases}
\pi-\arcsin\left(\frac{|x|\sin(|\angle x|)}{r_{*}}\right) & ,|x|\cos(\angle x)>|x^{*}|,\\
\arcsin\left(\frac{|x|\sin(|\angle x|)}{r_{*}}\right) & ,|x|\cos(\angle x)\leq|x^{*}|.
\end{cases}
\]
\end{lem}
\begin{IEEEproof}
The probability that the point $x\in\mathbb{R}^{2}$ belongs to $\mathcal{V}^{*}$
is equal to the probability that there does not exist a point of $\Phi_{a}\setminus\{x^{*}\}$
within the set $\mathcal{A}$, which equals $e^{-\lambda_{a}\pi|\mathcal{A}|}$
\cite{Baccelli}. For $|x|\leq|x^{*}|$ and $\angle x=0$, it is a
simple geometrical observation that $\mathcal{A}=\textrm{Ø}$, therefore,
$|\mathcal{A}|=0$. When $x^{*}=o$, $\mathcal{A}=\mathcal{B}(x,|x|)\setminus\mathcal{B}(o,0)=\mathcal{B}(x,|x|)$,
and, therefore, $|\mathcal{A}|=\pi|x|^{2}$. For all other cases,
$|\mathcal{A}|$ can be computed by the same approach as in the proof
of \cite[Theorem 1]{Handoff}. The procedure is straightforward but
tedious and is omitted.
\end{IEEEproof}
A simple lower bound for $p_{c}(x\mid x^{*})$ directly follows by
noting that $\mathcal{A}\subseteq\mathcal{B}(x,|x-x^{*}|)$ for all
$x\in\mathbb{R}^{2}$.
\begin{cor}
\label{cor: PCA bound}The PCA function is lower bounded as 
\begin{equation}
p_{c}(x\mid x^{*})\geq e^{-\lambda_{a}\pi|x-x^{*}|^{2}},x\in\mathbb{R}^{2},\label{eq:PCA lower bound}
\end{equation}

\noindent with equality if and only if $x^{*}=o$.
\end{cor}
\emph{Remark}: The right-hand side of (\ref{eq:PCA lower bound})
equals the probability that $x$ belongs to the Voronoi cell of the
AP positioned at $x^{*}$ when the latter is\emph{ not }conditioned
on being the closest AP to the origin or any other point in $\mathbb{R}^{2}$.

\begin{figure}
\centering{}\includegraphics[width=0.6\columnwidth]{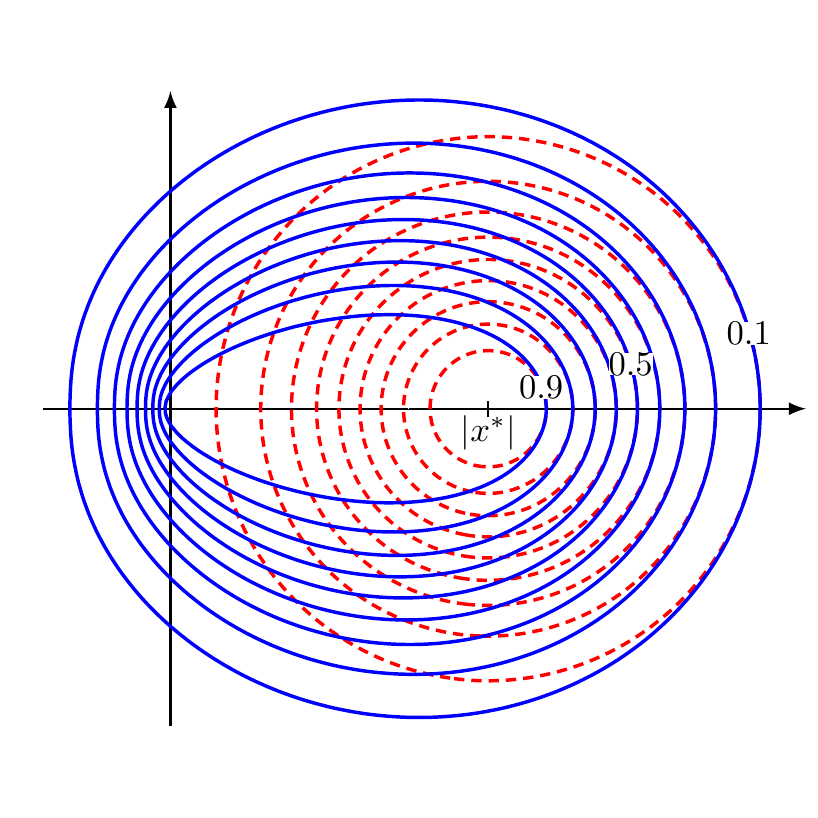}\caption{\label{fig:PCA}Contours of $p_{c}(x\mid x^{*})$ (solid lines), corresponding
to probabilities $0.1$ to $0.9$ in steps of $0.1$, for $|x^{*}|=1/\sqrt{\lambda_{a}}$.
The points of the line segment joining $o$ to $x^{*}$ are the only
ones with $p_{c}(x|x^{*})=1$. The corresponding contours of $e^{-\lambda_{a}\pi|x-x^{*}|^{2}}$
are also shown (dashed lines).}
\end{figure}

Figure \ref{fig:PCA} depicts $p_{c}(x\mid x^{*})$ for the case where
$|x^{*}|=1/\sqrt{\lambda_{a}}$ (behavior is similar for other values
of $|x^{*}|>0$). Note that, unless $x^{*}=o$, $p_{c}(x\mid x^{*})$
is not circularly symmetric w.r.t. any point in $\mathbb{R}^{2}$,
with its form suggesting that points isotropically distributed in
the vicinity of $x^{*}$ are more probable to lie within $\mathcal{V}^{*}$
than points isotropically distributed in the vicinity of $o$. 

\begin{figure}
\noindent \centering{}\includegraphics[width=1\columnwidth]{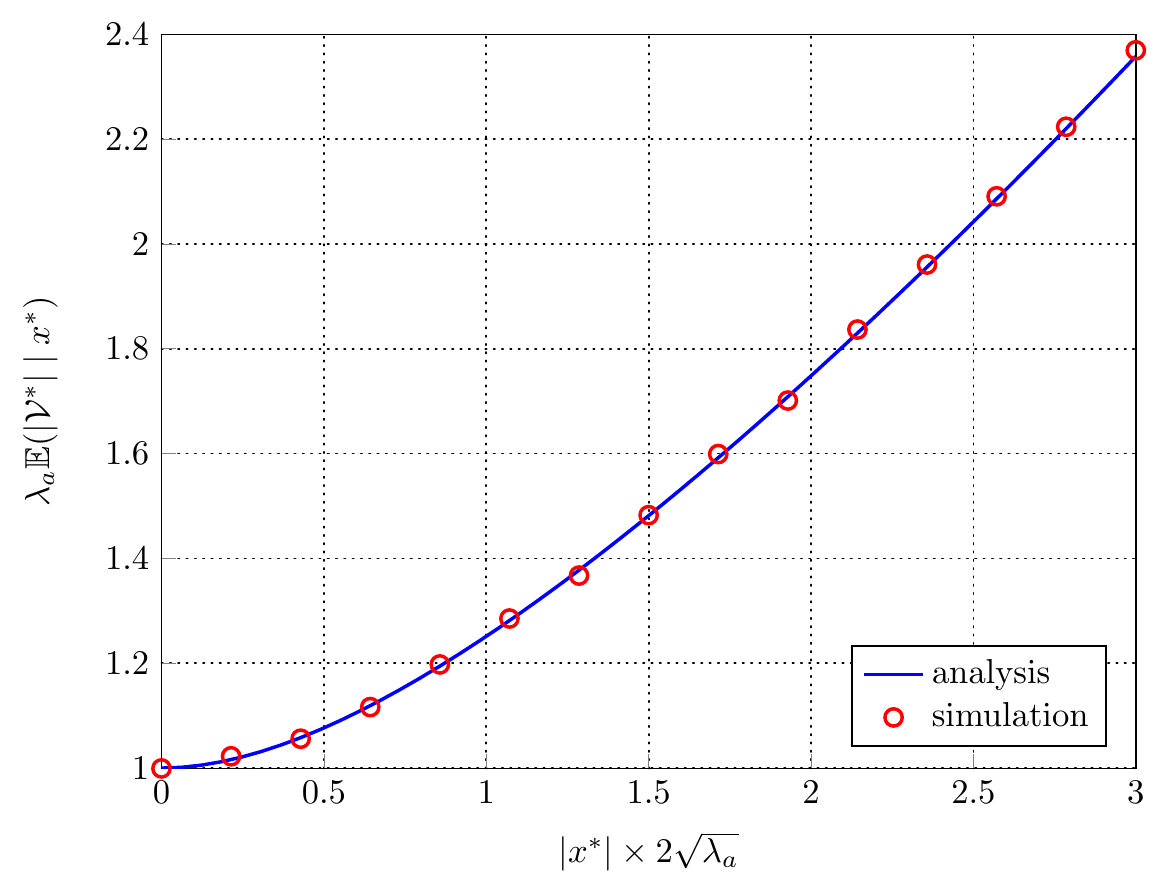}\caption{\label{fig:Average cell area}Average area of $\mathcal{V}^{*}$ conditioned
on $x^{*}$ as a function of the (normalized) distance of $x^{*}$
from the origin.}
\end{figure}

The lower bound $e^{-\lambda_{a}\pi|x-x^{*}|^{2}},x\in\mathbb{R}^{2}$,
is also shown in Fig. \ref{fig:PCA}, clearly indicating the \emph{probabilistic
expansion} \emph{effect} of the Voronoi cell of an AP, when the latter
is conditioned on being the closest to the origin. This cell expansion
effect is also demonstrated in Fig. \ref{fig:Average cell area} where
the conditional average cell area, equal to 
\begin{align}
\mathbb{E}(|\mathcal{V}^{*}|\mid x^{*}) & =\mathbb{E}\left(\left.\int_{\mathbb{R}^{2}}\mathbb{I}(x\in\mathcal{V}^{*})dx\right|x^{*}\right)\nonumber \\
 & =\int_{\mathbb{R}^{2}}p_{c}(x|x^{*})dx,\label{eq:average_cell_area}
\end{align}

\noindent is plotted as a function of the distance $|x^{*}|$, with
the integral of (\ref{eq:average_cell_area}) evaluated numerically
using Lemma \ref{lem:PCA}. Simulation results are also depicted serving
as a verification of the validity of Lemma \ref{lem:PCA}. It can
be seen that the average area of the guard region increases with $|x^{*}|$,
which implies a corresponding increase of the average number of UEs
that lie within the region. However, as will be discussed in the following,
even though resulting in more UEs silenced on average, an increasing
$|x^{*}|$ \emph{does not necessarily} imply improved protection from
UE interference, depending on the receiver position.

The interference statistics of the UE-generated interference power
are given in the following result.

\begin{prop}
\label{prop:Laplace type-II}The Laplace transform $\mathcal{L}_{I_{x_{R},u}}(s\mid x^{*})\triangleq\mathbb{E}(e^{-sI_{x_{R},u}}\mid x^{*})$
of the UE-generated interference power $I_{x_{R},u}$, conditioned
on $x^{*}$, is lower bounded as the right-hand side of (\ref{eq:radial Laplace transform})
with $P=P_{u}$, $\alpha=\alpha_{u}$ and $\lambda_{z}(r)=\lambda_{x_{R},u}(r),r\geq0$,
where
\begin{equation}
\lambda_{x_{R},u}(r)\triangleq\lambda_{u}\left(1-\frac{1}{2\pi}\int_{0}^{2\pi}p_{c}\left((r,\theta)+x_{R}\mid x^{*}\right)d\theta\right),\label{eq:UE density}
\end{equation}

\noindent with $p_{c}(x\mid x^{*})$ as given in Lemma \ref{lem:PCA}.
\end{prop}
\begin{IEEEproof}
See Appendix B.
\end{IEEEproof}
The following remarks can be made:
\begin{enumerate}
\item The statistics of the interfering UE point process when averaged over
$\mathcal{V}^{*}$ (equivalently, over $\Phi_{a}\setminus\{x^{*}\}$)
\emph{do not }correspond to a PPP, even though this is the case when
a fixed realization of $\mathcal{V}^{*}$ is considered. Therefore,
it cannot be expected that Lemma \ref{lem: Laplace Poisson theory}
applies for $\mathcal{L}_{I_{x_{R},u}}(s\mid x^{*})$. However, as
Prop. \ref{prop:Laplace type-II} shows, when the interfering point
process is \emph{treated} in the analysis as an PPP of an appropriately
defined equivalent density function $\lambda_{x_{R},u}(r)$, a tractable
lower bound for $\mathcal{L}_{I_{x_{R},u}}(s\mid x^{*})$ is obtained,
which will be shown to be tight in the numerical results section.
\item For $x^{*}\neq o$, the non circularly-symmetric form of the PCA function
results in an integral-form expression for the equivalent density
$\lambda_{x_{R},u}(r)$ that depends in general on $\angle x_{R}$,
in contrast to the case for $\lambda_{x_{R},a}(r)$ (see Sec. III.A,
Remark 2). A closed form expression for the equivalent density is
only available for $x^{*}=o$ discussed below.
\item When the receiver position is specified as $|x_{R}|\angle x_{R}$
with $\angle x_{R}$ uniformly distributed over $[-\pi,\pi)$, a straightforward
extension of the proof of Prop. \ref{prop:Laplace type-II} results
in the same lower bound for the conditioned Laplace transform of $I_{x_{R},u}$
as in Prop. \ref{prop:Laplace type-II}, with $\lambda_{x_{R},u}(r)$
replaced by 
\begin{equation}
\lambda_{|x_{R}|,u}(r)\triangleq\frac{1}{2\pi}\int_{-\pi}^{\pi}\lambda_{x_{R},u}(r)d\angle x_{R},\label{eq:average UE density}
\end{equation}
which is independent of $\angle x_{R}$.
\end{enumerate}
Although $\lambda_{x_{R},u}(r)$ is not available in closed form when
$x^{*}\neq o$, the numerical integration over a circular contour
required in (\ref{eq:UE density}) is very simple. With $\lambda_{x_{R},u}(r),r\geq0$,
pre-computed, evaluation of the lower bound for $\mathcal{L}_{I_{x_{R},u}}(s\mid x^{*})$
is of the same (small) numerical complexity as the evaluation of $\mathcal{L}_{I_{x_{R},a}}(s\mid x^{*})$.
Moreover, a closed-form upper bound for $\lambda_{x_{R},u}(r)$ is
available, which can in turn be used to obtain a looser lower bound
for $\mathcal{L}_{I_{x_{R},u}}(s\mid x^{*})$ that is independent
of $\angle x_{R}$. The bound is tight in the sense that it corresponds
to the exact equivalent density when $x^{*}=o$.
\begin{prop}
\label{prop:general bound}The equivalent density $\lambda_{x_{R},u}(r)$
of Prop. \ref{prop:Laplace type-II} is upper bounded as
\begin{equation}
\lambda_{x_{R},u}(r)\leq\lambda_{u}\left[1-e^{-\lambda_{a}\pi(|x^{*}|^{2}+|x_{R}|^{2}+r^{2})}\mathcal{I}_{0}(\lambda_{a}2\pi\bar{x}r)\right],\label{eq:general_upper_bound_for_density}
\end{equation}

\noindent where $\bar{x}\triangleq\max(|x^{*}|,|x_{R}|)$ and $\mathcal{I}_{0}(\cdot)$
denotes the zero-order modified Bessel function of the first kind.
Equality holds only when $x^{*}=o$.
\end{prop}
\begin{IEEEproof}
See Appendix C.
\end{IEEEproof}
An immediate observation of the above result is that, given $x_{R}$,
the equivalent interfering UE density decreases for any $x^{*}\neq o$
compared to the case $x^{*}=o$. This behavior is expected due to
the cell expansion effect described before. However, as the bound
of (\ref{eq:general_upper_bound_for_density}) is independent on $\angle x_{R}$,
it does not offer clear information on how the interferering UE density
changes when different values of $x^{*}\neq o$ are considered for
a given $x_{R}$. In order to obtain futher insights on the UE-generated
interference properties, $\lambda_{x_{R},u}(r)$ is investigated in
detail in the following section for certain special instances of $x_{R}$
and/or $x^{*}$, which are of particular interest in cellular networks.

\section{Analysis of Special Cases of UE-Generated Interference}

\subsection{$x_{R}=x^{*}$}

This case corresponds to a standard uplink cellular transmission with
nearest AP association, no intra-cell interference, and interference
generated by UEs outside $\mathcal{V}^{*}$ operating in uplink and/or
D2D mode. This case was previously investigated in \cite{UL Comm Lett,Andrews joint UL-DL}
for $\lambda_{u}=\lambda_{a}$, with heuristically introduced equivalent
densities averaged over $x^{*}$ (i.e., these densities are not expected
to be valid for an arbitrary value of $x^{*}$). In this section,
a more detailed investigation of the equivalent density properties
is provided.

For this case, a tighter upper bound than the one in Prop. \ref{prop:general bound}
is available for $\lambda_{x^{*},u}(r)$, which, interestingly, is
independent of $|x^{*}|$, in the practical case when $x^{*}\neq o$.
\begin{lem}
\label{lem:Upper_bound_of_UEs_to_AP_density}For $x_{R}=x^{*}$, the
equivalent density $\lambda_{x_{R},u}(r)=\lambda_{x^{*},u}(r)$ of
Prop. \ref{prop:Laplace type-II} is upper bounded as
\begin{equation}
\lambda_{x^{*},u}(r)\leq\lambda_{u}\left(1-e^{-\lambda_{a}\pi r^{2}}\right),\label{eq:UEs_to_AP_density_bound}
\end{equation}

\noindent with equality if and only if $x^{*}=o$.
\end{lem}
\begin{IEEEproof}
Follows by replacing the term $p_{c}\left((r,\theta)+x_{R}\mid x^{*}\right)=p_{c}\left((r,\theta)+x^{*}\mid x^{*}\right)$
appearing in (\ref{eq:UE density}) with its bound given in Cor. \ref{cor: PCA bound},
which evaluates to $e^{-\lambda_{a}\pi r^{2}}$.
\end{IEEEproof}
The bound of (\ref{eq:UEs_to_AP_density_bound}) indicates that $\lambda_{x^{*},u}(r)$
tends to zero at least as fast as $\mathcal{O}(r^{2})$ for $r\rightarrow0$,
\emph{irrespective of} $x^{*}$. The following exact statement on
the asymptotic behavior of $\lambda_{x^{*},u}(r)$ shows that $\lambda_{x^{*},u}(r)\sim cr^{2},r\rightarrow0$,
with the value of $c$ independent of $|x^{*}|$ when $x^{*}\neq o$.
\begin{prop}
\label{prop:UE_to_AP_density_asymptotics}For $x_{R}=x^{*}$, the
equivalent density $\lambda_{x_{R},u}(r)=\lambda_{x^{*},u}(r)$ of
Prop. \ref{prop:Laplace type-II} equals
\begin{equation}
\lambda_{x^{*},u}(r)=\lambda_{u}\lambda_{a}b\pi r^{2}+\mathcal{O}(r^{3}),r\rightarrow0,\label{eq:AP density asymptotic}
\end{equation}

\noindent with $b=1$ for $x^{*}=o$ and $b=1/2$ for $x^{*}\neq o$.
\end{prop}
\begin{IEEEproof}
For $x^{*}=o$, the result follows directly from Lemma \ref{lem:Upper_bound_of_UEs_to_AP_density}.
For $x^{*}\neq o$, it can be shown by algebraic manipulation based
on Lemma \ref{lem:PCA}, that the term $p_{c}\left((r,\theta)+x_{R}\mid x^{*}\right)=p_{c}\left((r,\theta)+x^{*}\mid x^{*}\right)$
appearing in (\ref{eq:UE density}) equals
\[
\lambda_{u}\lambda_{a}\left(q\pi+(-1)^{q}\arccos(\sin(\theta))+\cos(\theta)\sin(\theta)\right)r^{2}+\mathcal{O}(r^{3}),
\]

\noindent for $r\rightarrow0$, where $q=0$ for $\theta\in[-\pi,-\pi/2]\cup[\pi/2,\pi)$
and $q=1$ for $\theta\in(-\pi/2,\pi/2)$. Substituting this expression
in (\ref{eq:UE density}) and performing the integration leads to
the result.
\end{IEEEproof}
Equation (\ref{eq:AP density asymptotic}) indicates that imposing
a guard region $\mathcal{V}^{*}$ when $x^{*}\neq o$ reduces the
interfering UE density by half in the close vicinity of $x^{*}$,
compared to the case $x^{*}=o$. This effect is similar to the behavior
of $\lambda_{x^{*},a}(r)$ discussed in Sec. III. A. However, $\lambda_{x^{*},a}(0)=\lambda_{a}/2$
, whereas $\lambda_{x^{*},u}(0)=0$, clearly demonstrating the effectiveness
of the guard region for reducing UE-generated interference in the
vicinity of $x^{*}$.

\begin{figure}
\noindent \centering{}\includegraphics[width=1\columnwidth]{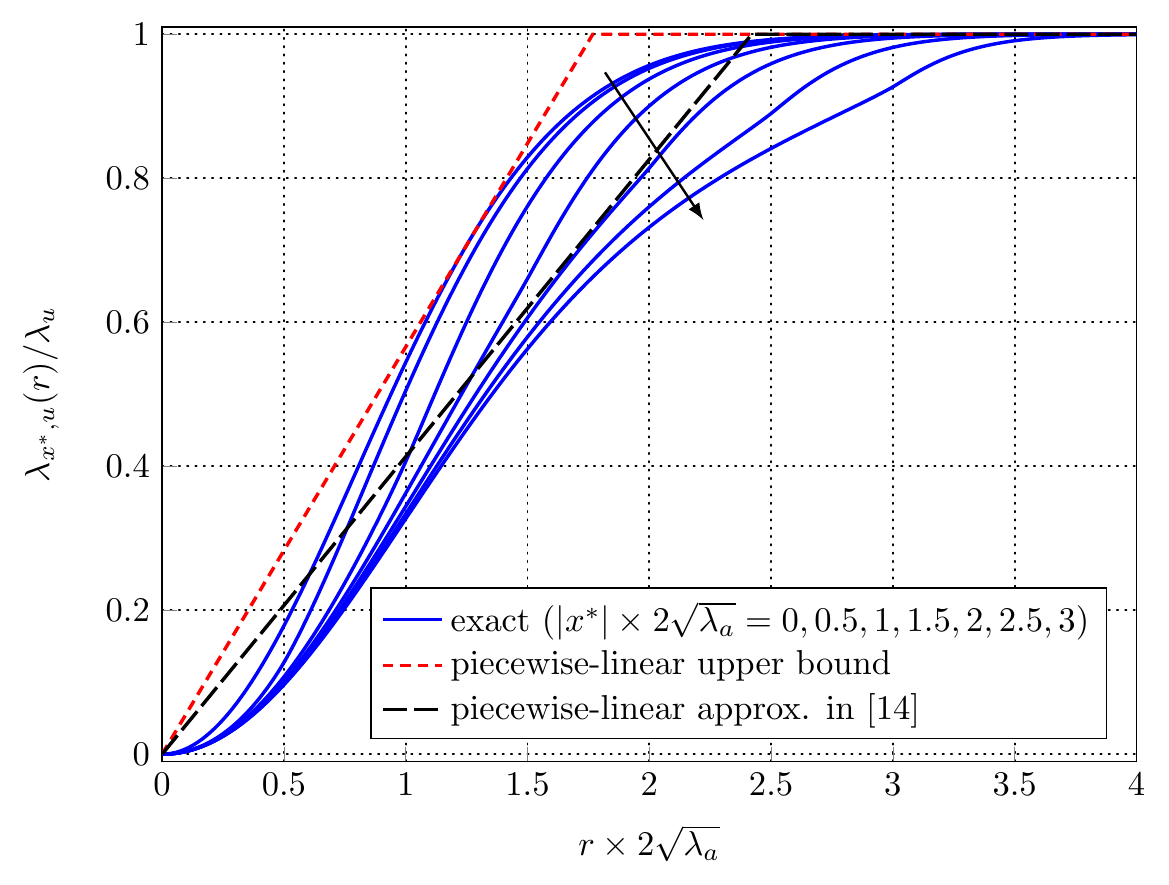}\caption{\label{fig:UEs_to_AP_density}Equivalent density function $\lambda_{x^{*},u}(r)$
for various $|x^{*}|$ (solid curves), with the direction of the arrow
corresponding to increasing $|x^{*}|$. A piecewise-linear approximation
and bound are also shown.}
\end{figure}

For non-asymptotic values of $r$, the behavior of $\lambda_{x^{*},u}(r)$
can only be examined by numerical evaluation of (\ref{eq:UE density}).
Figure \ref{fig:UEs_to_AP_density} depicts the normalized equivalent
density $\lambda_{x^{*},u}(r)/\lambda_{u}$ for various values of
$|x^{*}|$ in the range of practical interest $[0,3/(2\sqrt{\lambda_{a}})]$
(note that $\mathbb{P}(|x^{*}|>3/(2\sqrt{\lambda_{a}}))=e^{-\lambda_{a}\pi\left(3/(2\sqrt{\lambda_{a}})\right)^{2}}\approx8.5\times10^{-4})$.
It can be seen that for non-asymptotic values of $r$, $\lambda_{x^{*},u}(r)$
is a decreasing function of $|x^{*}|$ for all $r>0$, corresponding
to a reduced UE-generated interference. This is expected since, as
evident from Fig. \ref{fig:PCA}, the cell expansion effect with increasing
$|x^{*}|$ can only improve the interference protection at $x^{*}$.

Interestingly, the dependence of $\lambda_{x^{*},u}(r)$ on $|x^{*}|$,
although existing, can be seen to be rather small. This observation,
along with Prop. \ref{prop:UE_to_AP_density_asymptotics}, strongly
motivates the consideration of a single curve as an approximation
to the family of curves in Fig. \ref{fig:UEs_to_AP_density}. A natural
approach is to consider the upper bound of (\ref{eq:UEs_to_AP_density_bound}),
which has the benefit of being available in closed form. This is essentially
the approach that was heuristically employed in \cite{Andrews joint UL-DL}
without observing that it actually corresponds to a (tight) bound
for the equivalent interferer density.

Another approach is to consider a piecewise-linear curve, which, in
addition to having a simple closed form, results in a closed form
approximate expression for the bound of $\mathcal{L}_{I_{x^{*},u}}(s\mid x^{*})$
as given below (details are straightforward and omitted).
\begin{lem}
With the piecewise-linear approximation 
\begin{equation}
\lambda_{x^{*},u}(r)\approx\lambda_{u}\min\left(\delta r,1\right),r\geq0,\label{eq:piecewise-linear density}
\end{equation}

\noindent for some $\delta>0$, the lower bound for $\mathcal{L}_{I_{x^{*},u}}(s\mid x^{*})$
equals
\[
\exp\left\{ -\lambda_{u}\pi\left[C(P_{u}s)^{2/\alpha}+\frac{1}{\delta^{2}}\left(\frac{2}{3}\tilde{F}\left(3\right)-\tilde{F}\left(2\right)\right)\right]\right\} 
\]

\noindent where $C\triangleq(2\pi/\alpha)/\sin(2\pi/\alpha)$ and
$\tilde{F}(x)\triangleq{}_{2}F_{1}\left(1,\frac{x}{\alpha},1+\frac{x}{\alpha},\frac{-1}{sP_{u}\delta^{\alpha}}\right)$
with $_{2}F_{1}\left(\cdot\right)$ denoting the hypergeometric function.
\end{lem}
A piecewise-linear approximation as in (\ref{eq:piecewise-linear density})
was heuristically proposed in \cite{UL Comm Lett} for a slightly
more general system model than the one considered here. A closed form
formula for $\delta$ was provided, only justified as obtained by
means of curve fitting without providing any details on the procedure.
For the system model considered in this paper, this formula results
in $\delta\approx0.82687\sqrt{\lambda_{a}}$ and the corresponding
piecewise-linear approximation of the equivalent density is also depicted
in Fig. \ref{fig:UEs_to_AP_density}. It can be seen that this approximation
is essentially an attempt to capture the average behavior of $\lambda_{x^{*},u}(r)$
over $|x^{*}|$, thus explaining its good performance reported in
\cite{UL Comm Lett}. However, it is not clear why this particular
value of $\delta$ is more preferable than any other (slightly) different
value. A more rigorous approach for the selection of $\delta$ is
to consider the tightest piecewise-linear upper bound for $\lambda_{x^{*},u}(r)$,
which leads to another (looser) lower bound for $\mathcal{L}_{I_{x_{R},u}}(s\mid x^{*})$.
This bound corresponds to a value of $\delta\approx1.13118\sqrt{\lambda_{a}}$
, found by numerical optimization, and is also shown in Fig. \ref{fig:UEs_to_AP_density}.

\emph{Remark}: In \cite{Haenggi PVLP}, the case $x_{R}=x^{*}=o$
with VPLP distributed interferering UEs of density $\lambda_{u}=\lambda_{a}$
is considered. It is shown that the resulting interference properties,
averaged over $x^{*}$, are approximately similar to the properties
of the interference generated by PPP distributed interferers of an
equivalent density $\lambda_{a}(1-e^{-\frac{12}{5}\lambda_{a}\pi r^{2}})=\frac{12}{5}\lambda_{a}^{2}\pi r^{2}+\mathcal{O}(r^{3})$.
It is noted that, in the asymptotic ($r\rightarrow0$) regime, the
equivalent density of Prop. \ref{prop:UE_to_AP_density_asymptotics}
is smaller but has the same scaling order. This suggest that the analysis
of this section is a reasonable approximation also for the VPLP case.
This is verified numerically in Sec. V.

\subsection{$x^{*}=o$ }

This case corresponds to the typical node being an AP, i.e., a typical
AP. Note that, in contrast to the cases when $x^{*}\neq o$, the Voronoi
cell of the typical AP is \emph{not} conditioned to cover any point
in $\mathbb{R}^{2}$, therefore, no cell expansion effect is observed
and, for $x_{R}\neq o$, it is possible that the receiver lies outside
the guard region. Note that when it is the typical node/AP that is
transmitting to the receiver, the resulting operational scenario is
different from the standard downlink with nearest AP association \cite{Andrews tractable}
where the receiver lies within the Voronoi cell of the serving AP
by default. This scenario may arise, for example, in case of D2D communications,
where both communicating UEs receive data (e.g., control information)
from the same AP.

\begin{figure}
\noindent \centering{}\includegraphics[width=1\columnwidth]{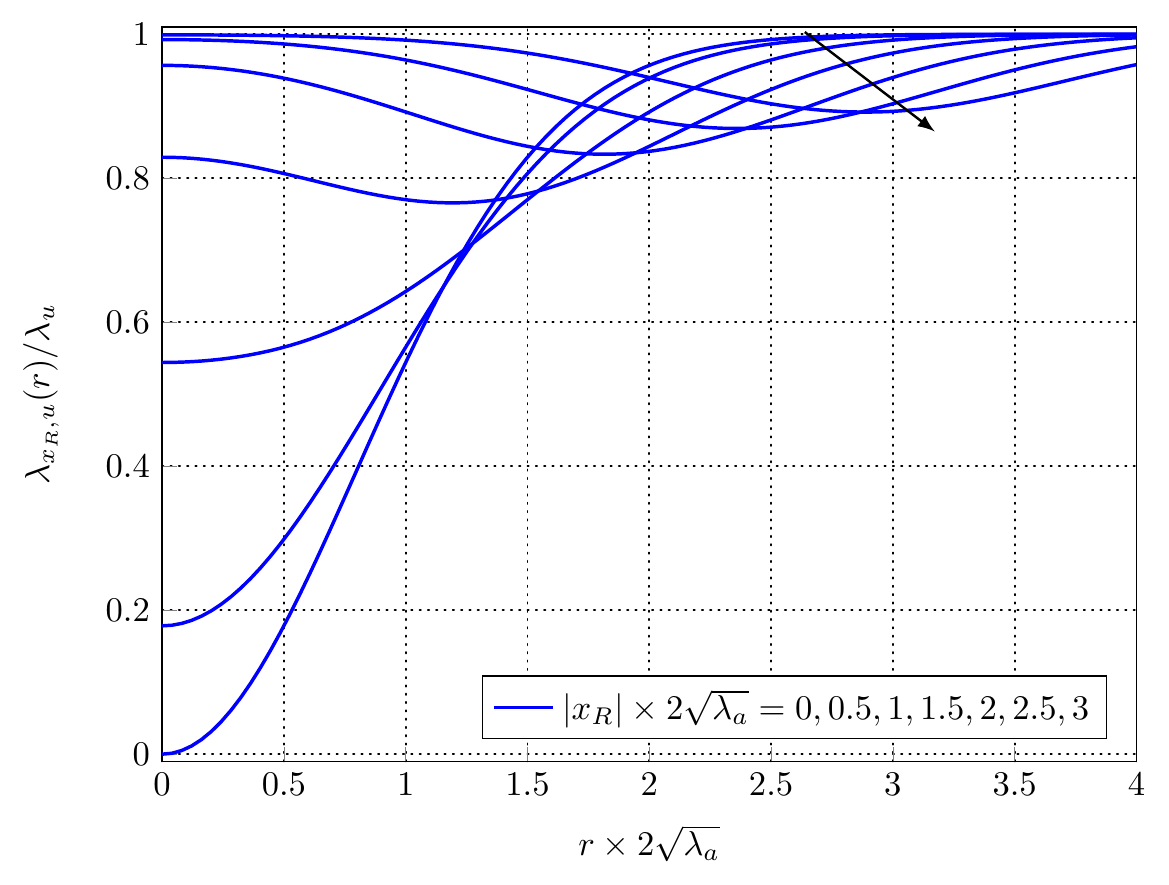}\caption{\label{fig:general_upper_bound_for_type_II} Equivalent density function
$\lambda_{x_{R},u}(r)$ when $x^{*}=o$ with the direction of the
arrow corresponding to increasing values of $|x_{R}|$.}
\end{figure}

The equivalent interfering UE density is given in Prop. \ref{prop:general bound}
and is depicted in Fig. \ref{fig:general_upper_bound_for_type_II}
for various values of $|x_{R}|$. Note that the case $x_{R}=o$ corresponds
to the typical AP in receive mode, whereas for $x_{R}\neq o$ the
typical AP transmits. It can be seen that increasing $|x_{R}|$ effectively
results in reduced protection in the close vicinity of the receiver
since the latter is more probable to lie near the edge or even outside
the Voronoi guard region.

\subsection{$x_{R}=o$ }

This case corresponds to the typical node being at receive mode. One
interesting scenario which corresponds to this case is downlink cellular
transmission with nearest AP association, where interference is due
to transmitting UEs (e.g., in uplink or D2D mode) that lie outside
the serving cell. 

When $x^{*}=o$, the setup matches the case $x_{R}=x^{*}$ with $x^{*}=o$,
discussed above, and it holds $\lambda_{o,u}(r|x^{*})=\lambda_{u}\left(1-e^{-\lambda_{a}\pi r^{2}}\right)$.
For $x^{*}\neq o$, the asymptotic behavior of $\lambda_{o,u}(r|x^{*})$
can be obtained by the same approach as in the proof of Prop. \ref{prop:UE_to_AP_density_asymptotics}.
\begin{prop}
\label{prop:UE_to_UE_density_asymptotics}For $x^{*}\neq o$ the equivalent
density function $\lambda_{x_{R},u}(r)=\lambda_{o,u}(r|x^{*})$ of
Prop. \ref{prop:Laplace type-II} equals
\[
\lambda_{o,u}(r)=\lambda_{u}\lambda_{a}\frac{8}{\pi}|x^{*}|r+\mathcal{O}(r^{2}),r\rightarrow0.
\]
\end{prop}
In stark contrast to the behavior of $\lambda_{x^{*},u}(r)$, $\lambda_{o,u}(r)$
tends to $0$ only linearly (instead of quadradically) as $r\rightarrow0$
when $x^{*}\neq o$, and with a rate that is also proportional to
(instead of independent of) $|x^{*}|$. This implies that the typical
node is not as well protected from interfering UEs as its nearest
AP, experiencing an increased equivalent interferer density in its
close vicinity as $|x^{*}|$ increases. This can be understood by
noting that with increasing $|x^{*}|$, the origin, although contained
in a guard region of (statistically) increasing area, is more probable
to lie near the cell edge (Fig. \ref{fig:system model} demonstrates
such a case.)

\begin{figure}
\noindent \centering{}\includegraphics[width=1\columnwidth]{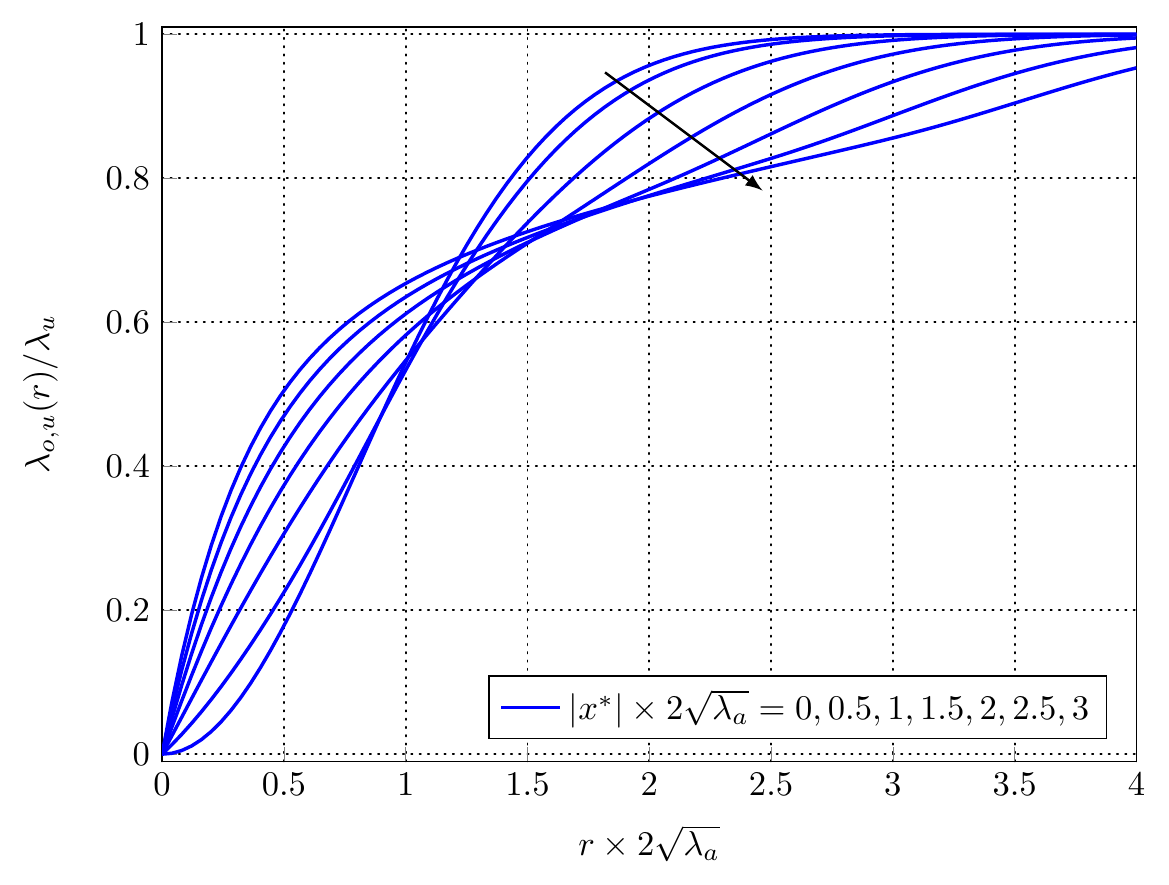}\caption{\label{fig:UE_to_UE_density}Equivalent density function $\lambda_{o,u}(r)$
for various $|x^{*}|$, with the direction of the arrow corresponding
to increasing $|x^{*}|$. }
\end{figure}

The strong dependence of $\lambda_{o,u}(r)$ on $|x^{*}|$ observed
for $r\rightarrow0$ also holds for all values of $r$. This is shown
in Fig. \ref{fig:UE_to_UE_density} where the numerically evaluated
$\lambda_{o,u}(r)$ is depicted for the same values of $|x^{*}|$
as those considered in Fig. \ref{fig:UEs_to_AP_density}. It can be
seen that $\lambda_{o,u}(r)$ increases with $|x^{*}|$ for all $r$
up to approximately $1-1.5$ times the average value of $|x^{*}|$
(equal to $1/(2\sqrt{\lambda_{a}})$), whereas it decreases with $|x^{*}|$
for greater $r$ .

This behavior of $\lambda_{o,u}(r)$, especially for $r\rightarrow0$,
does not permit the use of single function to be used as a reasonable
approximation or bound for $\lambda_{o,u}(r)$, irrespective of $|x^{*}|$,
as was done for the case for $\lambda_{x^{*},u}(r)$. Noting that
the upper bound of (\ref{eq:general_upper_bound_for_density}) has
the same value whenever $|x^{*}|$ or $|x_{R}|$ are zero, it follows
that the curves shown in Fig. \ref{fig:general_upper_bound_for_type_II}
with $|x_{R}|$ replaced by $|x^{*}|$ are upper bounds for the corresponding
curves of Fig. \ref{fig:UE_to_UE_density}. Unfortunately, it can
be seen that the bound is very loose for small $r$ and/or large $|x^{*}|$. 

\emph{Remark}: In \cite{Haenggi PVLP}, the case $x_{R}=o$ with VPLP
distributed interferering UEs of density $\lambda_{u}=\lambda_{a}$
is considered. It is shown that the resulting interference properties,
averaged over $x^{*}$, are approximately similar to the properties
of the interference generated by PPP distributed interferers of (equivalent)
density $\lambda_{a}(1-e^{-\frac{9}{4}\sqrt{\lambda_{a}}\pi r}+\frac{1}{2}r^{2}e^{-\frac{5}{4}\lambda r^{2}})=\frac{9}{4}\lambda_{a}^{3/2}\pi r+\mathcal{O}(r^{2})$.
As an approximation of this case, the equivalent density of Prop.
\ref{prop:UE_to_UE_density_asymptotics} can be averaged over $|x^{*}|$
resulting in $\frac{4}{\pi}\lambda_{a}^{3/2}\pi r+\mathcal{O}(r^{2})$.
It can be seen that the approximation is able to capture the asymptotic
scaling of the actual density, even though with a slightly smaller
factor, suggesting the validity of the analysis as a reasonable approximation
for this case as well. 

\begin{figure*}[t]
\subfloat[\label{fig:SIR_2_l1}]{\noindent \centering{}\includegraphics[width=1\columnwidth]{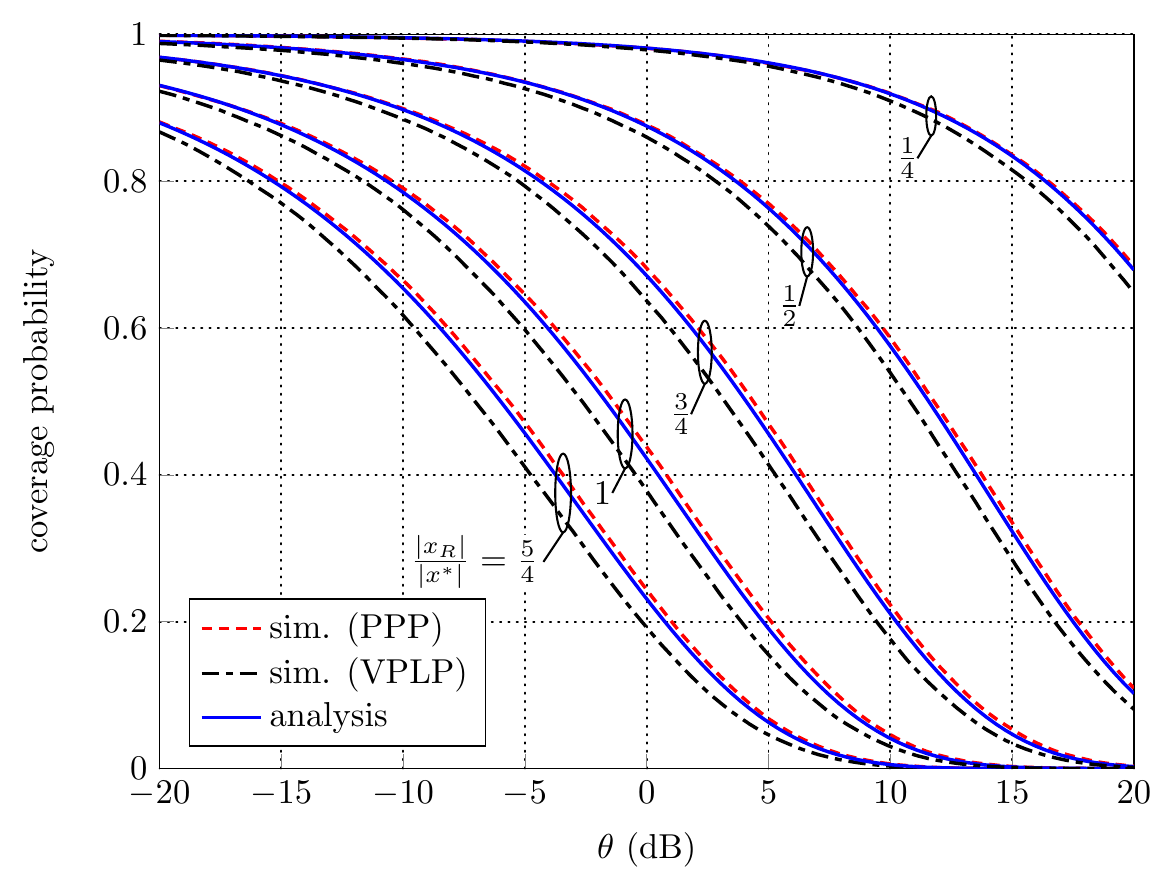}}\hfill{}\subfloat[\label{fig:SIR_2_l10}]{\noindent \begin{centering}
\includegraphics[width=1\columnwidth]{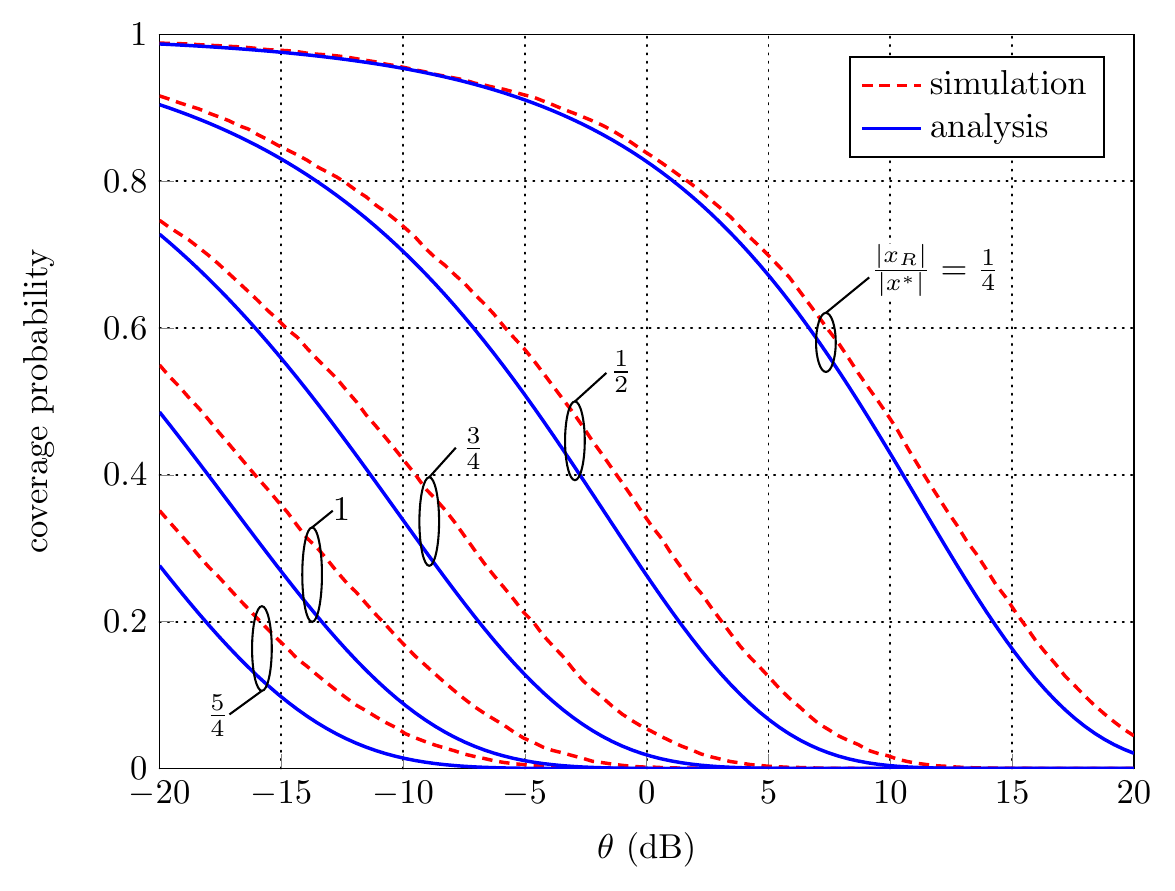}
\par\end{centering}
}\caption{\label{fig:SIR_2}Coverage probability of the link between the typical
node and an isotropically positioned receiver under UE interference
for (a) $\lambda_{u}=\lambda_{a}$, (b) $\lambda_{u}=10\lambda_{a}$. }
\end{figure*}

\subsection{$|x_{R}|\in(0,|x^{*}|),\angle x_{R}=0$}

The previous cases demonstrate how different the properties of the
UE-generated interference become when different receiver positions
are considered. This observation motivates the question of which receiver
position is best protected from UE interference given the positions
$o$ and $x^{*}\neq o$ of the typical node and nearest AP, respectively.
This question is relevant in, e.g., relay-aided cellular networks
where the communication between a UE and its nearest AP is aided by
another node (e.g., an inactive UE) \cite{Elkotby relay}. Although
the performance of a relay-aided communication depends on multiple
factors including the distances among the nodes involved and considered
transmission scheme \cite{Gamal}, a reasonable choice for the relay
position is the one experiencing less interference. 

By symmetry of the system geometry, this position should lie on the
line segment joining the origin with $x^{*}$ (inclusive). However,
as can be seen from Figs. \ref{fig:UEs_to_AP_density} and \ref{fig:UE_to_UE_density},
the shape of the equivalent interferer density does not allow for
a natural ordering of the different values of $x_{R}$. For example,
$\lambda_{x^{*},u}(r)$ is smaller than $\lambda_{o,u}(r)$ for small
$r$, whereas the converse holds for large $r$. Noting that the receiver
performance is mostly affected by nearby generated interference \cite{Haenggi=000026Ganti},
a natural criterion to order the interfering UE densities for varying
$x_{R}$ is their asymptotic behavior as $r\rightarrow0$. For $|x_{R}|=|x^{*}|$
and $|x_{R}|=o$, this is given in Props. \ref{prop:UE_to_AP_density_asymptotics}
and \ref{prop:UE_to_UE_density_asymptotics}, respectively. For all
other positions the asymptotic density is given by the following result. 
\begin{prop}
For $|x_{R}|\in(0,|x^{*}|),\angle x_{R}=0$, and $x^{*}\neq o$, the
equivalent density function $\lambda_{x_{R},u}(r)$ of Prop. \ref{prop:Laplace type-II}
equals
\begin{equation}
\lambda_{x_{R},u}(r)=\lambda_{u}\lambda_{a}\frac{8}{9\pi}\frac{(|x^{*}|/|x_{R}|)^{2}}{|x^{*}|-|x_{R}|}r^{3}+\mathcal{O}(r^{4}),r\rightarrow0.\label{eq:RL_density}
\end{equation}
\end{prop}
It follows that the optimal receiver position must have $|x_{R}|\in(0,|x^{*}|)$
since, in this case, the density scales as $\mathcal{O}(r^{3})$ instead
of $\mathcal{O}(r)$ and $\mathcal{O}(r^{2})$ when $|x_{R}|=0$ and
$|x_{R}|=|x^{*}|$, respectively. Its value can be easily obtained
by minimization of the expression in (\ref{eq:RL_density}) w.r.t.
$|x_{R}|$.
\begin{cor}
The receiver position that experiences the smallest equivalent UE
interferer density in its close vicinity lies on the line segment
joining the origin to $x^{*}$ and is at distance $|x^{*}|/\sqrt{2}$
from the origin. The corresponding equivalent density equals $\lambda_{x_{R},u}(r)=\lambda_{u}\lambda_{a}\frac{32}{9\pi}r^{3}+\mathcal{O}(r^{4}),r\rightarrow0$.
\end{cor}

\section{Numerical Examples}

In this section, numerical examples will be given, demonstrating the
performance of various operational scenarios in cellular networks
that can be modeled as special cases of the analysis presented in
the previous sections. The system performance metric that will be
considered is $\mathcal{L}_{I_{x_{R},k}}(\rho^{\alpha_{k}}\theta|x^{*})$,
with $k\in\{a,u\}$ depending on whether AP- or UE-generated interference
is considered, with $\rho>0$, $\theta>0$ given. Note that this metric
corresponds to the\emph{ coverage probability} that the signal-to-interference
ratio (SIR) at $x_{R}$ is greater than $\theta$, when the distance
between transmitter and receiver is $\rho$, the direct link experiences
a path loss exponent $a_{k}$ and Rayleigh fading, and the transmit
power is equal to $P_{k}$ \cite{Haenggi=000026Ganti}. For simplicity
and w.l.o.g., all the following examples correspond to $\lambda_{a}=1$
and $\alpha_{a}=\alpha_{u}=4$.

The analytical results of the previous sections will be compared against
simulated system performance where, in a single experiment, the distribution
of interfering APs and UEs is generated as follows. Given the AP position
closest to the origin, $x^{*}$, the interfering AP positions are
obtained as a realization of a PPP of density as in (\ref{eq:andrews_density}).
Afterwards, an independent realization of a HPPP of density $\lambda_{u}$
is generated, which, after removing the points lying within $\mathcal{V}^{*}$,
represents the interferering UE positions. 

In addition, for cases where the UE-generated interference is of interest,
a VPLP process of density $\lambda_{u}=\lambda_{a}$ is also simulated,
corresponding to scenarios where one UE from each Voronoi cell in
the system transmits (as in the standard uplink scenario with nearest
AP association).

\subsection{D2D Link not Necessarily Contained in a Single Cell}

In this example, UE-generated inteference is considered at a receiver
position $x_{R}\neq o$ with $\angle x_{R}$ uniformly distributed
in $[-\pi,\pi)$ and $\rho=|x_{R}|$. This case may model a D2D link
where the typical node is a UE that directly transmits to a receiver
isotropically distributed at a distance $|x_{R}|>0$ . A guard region
$\mathcal{V}^{*}$ is established by the closest AP to the typical
node, however, it is not guaranteed to include $x_{R}$, i.e., the
D2D nodes may not be contained within the same cell, which is a common
scenario that arises in practice, referred to as cross-cell D2D communication
\cite{cross-cell D2D}. 

With interference generated from other UEs in D2D and/or uplink mode,
a lower bound for $\mathcal{L}_{I_{x_{R},u}}(\rho^{\alpha_{u}}\theta|x^{*})$
can be obtained using Prop. \ref{prop:Laplace type-II} with an equivalent
density function $\lambda_{|x_{R}|,u}(r)$ as given in (\ref{eq:average UE density}).
Figure \ref{fig:SIR_2} shows this lower bound for $\lambda_{u}=\lambda_{a}$
(Fig. \ref{fig:SIR_2_l1}) and $\lambda_{u}=10\lambda_{a}$ (Fig.
\ref{fig:SIR_2_l10}). The position $x^{*}$ is assumed to be isotropically
distributed with $|x^{*}|=1/(2\sqrt{\lambda_{a}})$ \footnote{Recall that this value is the average distance of the closest AP from
the typical node.} and various values of the ratio $|x_{R}|/|x^{*}|$ are considered.
It can be seen that, compared to the simulation under the PPP interference
assumption, the quality of the analytical lower bound depends on $\lambda_{u}$.
For $\lambda_{u}=\lambda_{a}$, it is very close to the exact coverage
probability, whereas for $\lambda_{u}=10\lambda_{a}$, it is reasonably
tight and able to capture the behavior of the exact coverage probability
over varying $|x_{R}|$. Also, for $\lambda_{u}=\lambda_{a}$ and
a VPLP interference, the analytical expression provides a reasonably
accurate approximation for the performance of this analytically intractable
case. As expected, performance degradation is observed with increasing
$|x_{R}|$, due to both increasing path loss of the direct link as
well as increased probability of the receiver lying outside $\mathcal{V}^{*}$. 

\subsection{AP-to-D2D-Receiver Link}

In this example, AP-generated inteference is considered at a receiver
position $x_{R}\neq o$ with $\angle x_{R}$ uniformly distributed
in $[-\pi,\pi)$ and $\rho=|x^{*}-x_{R}|$. This case is similar to
the previous, however, considering AP-generated interference and the
receiver at $x_{R}$ receiving data from the AP node at $x^{*}$.
This case models the scenario where the typical node establishes a
D2D connection with a node at $x_{R}$, and the AP at $x^{*}$ sends
data to $x_{R}$ via a dedicated cellular link (e.g., for control
purposes or for implementing a cooperative transmission scheme). Note
that, in contrast to the previous case, the link distance $\rho$
is random and equal to $\rho=|x^{*}|+|x_{R}|-2|x^{*}||x_{R}|\cos\psi$,
with $\psi$ uniformly distributed in $[-\pi,\pi)$. The exact coverage
probability of this link can be obtained using Prop. \ref{prop:type-I Laplace}
followed by a numerically computed expectation over $\psi$.

\begin{figure}[t]
\noindent \centering{}\includegraphics[width=1\columnwidth]{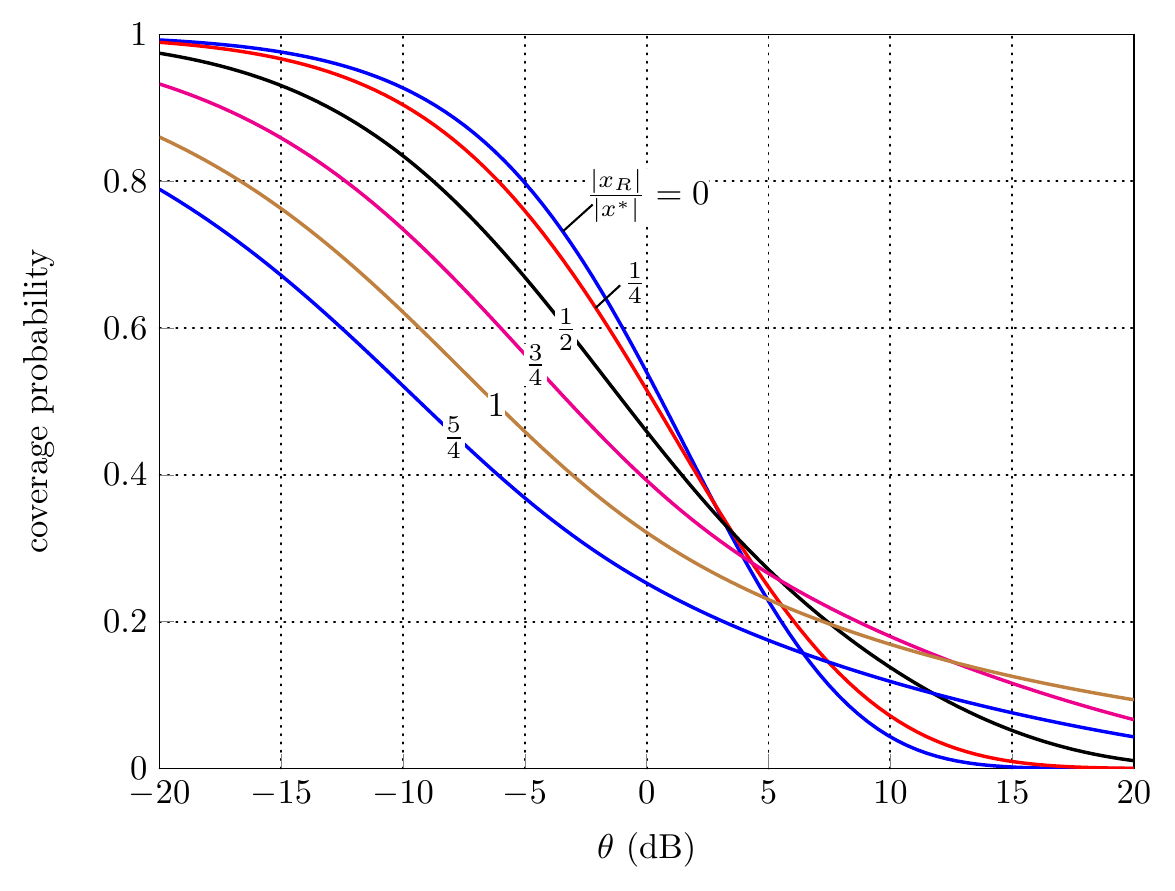}\caption{\label{fig:SIR_1}Coverage probability for the link between the AP
positioned at $x^{*}$ ($|x^{*}|=1/(2\sqrt{\lambda_{a}})$) and an
isotropicaly positioned receiver.}
\end{figure}

Figure \ref{fig:SIR_1} shows $\mathbb{E}(\mathcal{L}_{I_{x_{R},a}}(\rho^{\alpha_{a}}\theta|x^{*}))$
for an isotropically distributed $x^{*}$ with $|x^{*}|=1/(2\sqrt{\lambda_{a}})$,
and for various values of the ratio $|x_{R}|/|x^{*}|$. Note that
the case $|x_{R}|=0$ corresponds to the standard downlink transmission
model with nearest AP association \cite{Andrews tractable}. Monte
Carlo evaluation of the coverage probability (not shown here) perfectly
matches the analytical curves. It can be seen that increasing $|x_{R}|$
reduces the coverage probability for small SIR but increases it for
high SIR. This is in direct proportion to the behavior of the equivalent
interferer density $\lambda_{x_{R},a}(r)$ with increasing $|x_{R}|$
shown in Fig. \ref{fig:AP_to_RL_density_vs_rho_RL}.

\subsection{Uplink with Nearest AP Association}

\begin{figure}
\noindent \centering{}\includegraphics[width=1\columnwidth]{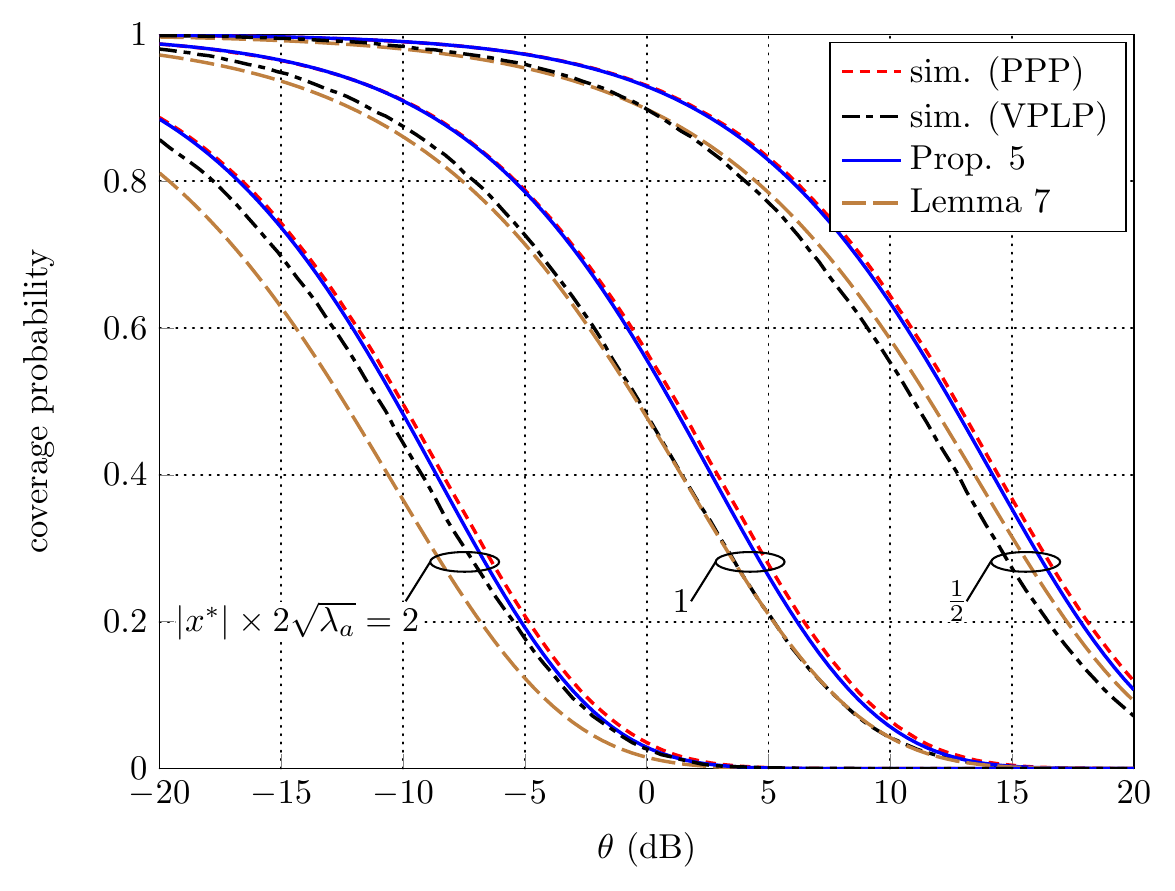}\caption{\label{fig:uplink SIR}Coverage probability of the uplink between
the typical node its nearest AP ($\lambda_{u}=\lambda_{a}$).}
\end{figure}

In this example, UE-generated interference is considered at a receiver
position $x_{R}=x^{*}$ with $\rho=|x^{*}|$ and $\lambda_{u}=\lambda_{a}$
corresponding to the conventional uplink transmission scenario with
nearest AP association and one active UE per cell (no cross-mode interference).
Figure \ref{fig:uplink SIR} shows the lower bound of $\mathcal{L}_{I_{x_{R},u}}(\rho^{\alpha_{u}}\theta|x^{*})$
obtained by Prop. \ref{prop:Laplace type-II} as well as the looser,
but more easily computable, lower bound obtained using the closed
form expression given in Lemma \ref{lem:Upper_bound_of_UEs_to_AP_density}
with $\delta=1.13118\sqrt{\lambda_{a}}$ (resulting in the tightest
bound possible with a piecewise-linear equivalent density function).
The coverage probability is computed for an AP at a distance $|x^{*}|$=$c/(2\sqrt{\lambda_{a}})$
from the typical node with $c=1/2,1,2$, roughly corresponding to
a small, average, and large uplink distance (performance is independent
of $\angle x^{*}$). 

As was the case in Fig. \ref{fig:SIR_2_l1}, Prop. \ref{prop:Laplace type-II}
provides a very tight lower bound for the actual coverage probability
(under the PPP model for the interfering UE positions). The bound
of Lemma \ref{lem:Upper_bound_of_UEs_to_AP_density}, although looser,
is nevertheless a reasonable approximation of the performance, especially
for smaller values of $|x^{*}|$. Compared to the VPLP model, the
PPP model provides an optimistic performance prediction that is, however,
reasonably tight, especially for large $|x^{*}|$. This observation
motivates its usage as a tractable approximation of the actual inteference
statistics as was also reported in \cite{UL Comm Lett,Andrews joint UL-DL}.
Interestingly, the bound of Lemma \ref{lem:Upper_bound_of_UEs_to_AP_density}
happens to provide an even better approximation for the VPLP performance
for $|x^{*}|$ close to or smaller than $1/(2\sqrt{\lambda_{a}})$. 

\subsection{Effect of Guard Region on UE-Generated Interference Protection}

\begin{figure}
\noindent \centering{}\includegraphics[width=1\columnwidth]{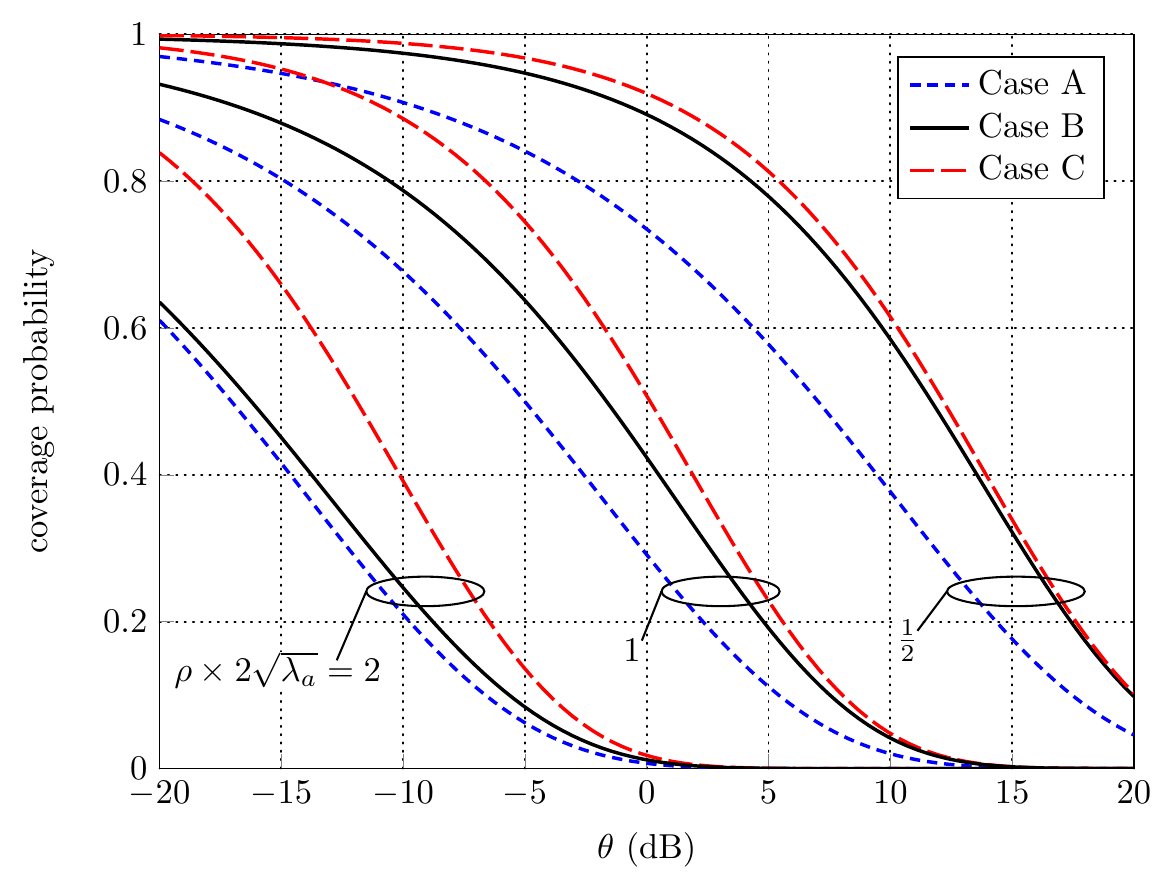}\caption{\label{fig:SIR_cases}Coverage probability under UE interference for
the links corresponding to the cases described in Sec. V. D.}
\end{figure}

In order to see the effectiveness of a Voronoi guard region for enhancing
link quality under UE-generated interference, the performance under
the following operational cases is examined.
\begin{itemize}
\item Case A (no guard region is imposed): This results in the standard
transmission model under an HPPP of interferers positions of density
$\lambda_{u}$ \cite{Haenggi=000026Ganti}. The exact coverage probability
is well known (see, e.g., \cite[Eq. 3.29]{Haenggi=000026Ganti}).
\item Case B (transmitter imposes a Voronoi guard region): This case corresponds
to $x^{*}=o$, $x_{R}\neq o$, modeling an non-standard downlink transmission
(see also description in Table \ref{tab:Special-Cases-of}).
\item Case C (receiver imposes a Voronoi guard region): This case corresponds
to $x_{R}=x^{*}=o$ modeling an non-standard uplink transmission (see
also description in Table \ref{tab:Special-Cases-of})
\end{itemize}
Cases B and C correspond to the analysis considered in Sec. V. B.
Assuming the same link distance $\rho$ for all cases, Fig. \ref{fig:SIR_cases}
shows the analytically obtained coverage probability (exact for Case
A, lower bound for Cases B and C), assuming $\lambda_{u}=\lambda_{a}$.
It can be seen that imposing a Voronoi guard region (Cases B and C)
is always beneficial, as expected. However, for large link distances,
Case B provides only marginal gain as the receiver is very likely
to be located at the edge or even outside the guard region. In contrast,
the receiver is always guaranteed to be protected under case C, resulting
in the best performance and significant gains for large link distances.

\section{Conclusion}

This paper considered the analytical characterization of cross-mode
inter-cell interference experienced in future cellular networks. By
employing a stochastic geometry framework, tractable expressions for
the interference statistics were obtained that are exact in the case
of AP-generated interference and serve as a tight lower bound in the
case of UE-generated interference. These expressions are based on
appropriately defined equivalent interferer densities, which provide
an intuitive quantity for obtaining insights on how the interference
properties change according to the type of interference and receiver
position. The considered system model and analysis are general enough
to capture many operational scenarios of cellular networks, including
conventional downlink/uplink transmissions with nearest AP association
as well as D2D transmissions between UEs that do not necessarily lie
in the same cell. The analytical expressions of this paper can be
used for sophisticated design of critical system aspects such as mode
selection and resource allocation towards getting the most out D2D-
and/or FD-enabled cellular communications. Interesting topics for
future research is investigation of the joint properties of AP- and
UE-generated interference, consideration of different transmit powers
for uplink and D2D UE transmissions, as well as extension of the analysis
to the case of heterogeneous, multi-tier networks.

\appendices{}

\section{Proof of Proposition \ref{prop:type-I Laplace}}

\begin{figure}
\noindent \begin{centering}
\includegraphics[width=0.7\columnwidth]{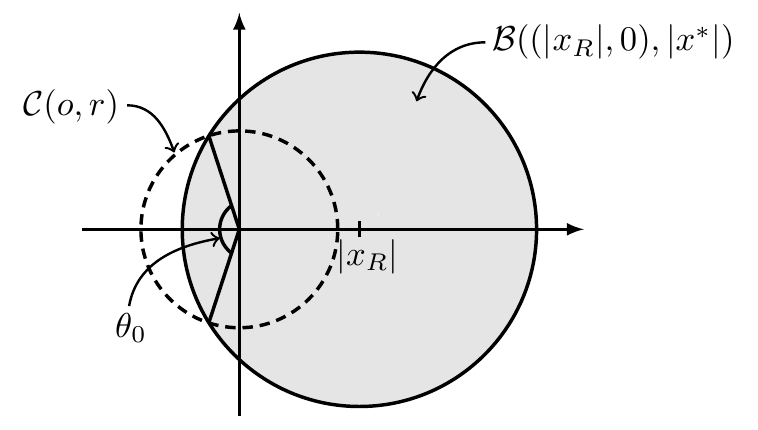}
\par\end{centering}
\caption{\label{fig:AP_to_RL_density_proof} Geometrical figure depicting the
angle $\theta_{0}$, which is equal to quantity $\int_{0}^{2\pi}\mathbb{I}\left((r,\theta)\protect\notin\mathcal{B}((|x_{R}|,0),|x^{*}|)\right)d\theta$
appearing in the proof of Prop. \ref{prop:type-I Laplace}.}
\end{figure}

Let $\tilde{\Phi}_{a}\triangleq\Phi_{a}\setminus\{x^{*}\}$ denote
the interfering AP point process. Given $x^{*}$, $\tilde{\Phi}_{a}$
is an inhomogeneous PPP of density $\tilde{\lambda}_{a}(x),x\in\mathbb{R}^{2}$,
defined in (\ref{eq:andrews_density}). Density $\tilde{\lambda}_{a}(x)$
is circularly symmetric w.r.t. $x_{R}=o$, which, using Lemma \ref{lem: Laplace Poisson theory},
directly leads to the Laplace transform expression of (\ref{eq:radial Laplace transform})
with radial density as in (\ref{eq:DL AP density}). Considering the
case $x_{R}\neq o$, it directly follows from Lemma \ref{lem: Laplace Poisson theory}
and (\ref{eq:2D integral Laplace transform}) that
\begin{align*}
\mathcal{L}_{I_{x_{R},a}}(s) & =\exp\left\{ -\int_{\mathbb{R}^{2}}\tilde{\lambda}_{a}(x+x_{R})\gamma(sP_{a}|x|^{-\alpha_{a}})dx\right\} 
\end{align*}

\noindent by a change of integration variable. By switching to polar
coordinates (centered at $o$) for the integration, the right-hand
side of (\ref{eq:radial Laplace transform}) results with $P=P_{a}$,
$\alpha=\alpha_{a}$ and $\lambda_{x_{R}}(r)=\lambda_{x_{R},a}(r)$,
where 
\begin{align*}
\lambda_{x_{R},a}(r) & =\frac{1}{2\pi}\int_{0}^{2\pi}\tilde{\lambda}_{a}((r,\theta)+x_{R})d\theta\\
 & =\frac{\lambda_{a}}{2\pi}\int_{0}^{2\pi}\mathbb{I}\left\{ (r,\theta)+x_{R}\notin\mathcal{B}(o,|x^{*}|)\right\} d\theta\\
 & =\frac{\lambda_{a}}{2\pi}\int_{0}^{2\pi}\mathbb{I}\left\{ (r,\theta)\notin\mathcal{B}(-x_{R},|x^{*}|)\right\} d\theta\\
 & =\frac{\lambda_{a}}{2\pi}\int_{0}^{2\pi}\mathbb{I}\left\{ (r,\theta)\notin\mathcal{B}((|x_{R}|,0),|x^{*}|)\right\} d\theta,
\end{align*}

\noindent with $x_{R}$ representing the polar coordinates of the
receiver in the first equality, with a slight abuse of notation. The
last equality follows by noting that the integral does not depend
on the value of $\angle x_{R}$. Evaluation of the final integral
is essentially the evaluation of an angle (see Fig. \ref{fig:AP_to_RL_density_proof}),
which can be easily obtained by elementary Euclidean geometry, resulting
in the expression of (\ref{eq:AP-interferer_density}).

\noindent 

\section{Proof of Proposition \ref{prop:Laplace type-II}}

It holds 
\begin{align*}
 & \mathcal{L}_{I_{x_{R},u}}(s\mid x^{*})\\
= & \mathbb{E}\left(\mathbb{E}\left(e^{-sI_{x_{R},u}}\mid\Phi_{a}\right)\mid x^{*}\right)\\
\stackrel{(a)}{=} & \mathbb{E}\left(\left.\exp\left\{ -\int_{\mathbb{R}^{2}}\tilde{\lambda}_{u}(x|\Phi_{a})\gamma(sP_{u}|x-x_{R}|^{-\alpha_{u}})dx\right\} \right|x^{*}\right)\\
\geq & \exp\left\{ -\int_{\mathbb{R}^{2}}\mathbb{E}\left(\tilde{\lambda}_{u}(x+x_{R}|\Phi_{a})\mid x^{*}\right)\gamma(sP_{u}|x|^{-\alpha_{u}})dx\right\} ,
\end{align*}

\noindent where $(a)$ follows by noting that, conditioned on $\Phi_{a}$,
the interfering UE point process equals $\Phi_{u}\setminus\mathcal{V}^{*}$,
which is distributed as an inhomogeneous PPP of density $\tilde{\lambda}_{u}(x|\Phi_{1}),x\in\mathbb{R}^{2}$
as given in (\ref{eq:UE density general}), and using Lemma \ref{lem: Laplace Poisson theory}.
The inequality is an application of Jensen's inequality with a change
of integration variable. Result follows by noting that $\mathbb{E}\left(\tilde{\lambda}_{u}(x|\Phi_{a})\mid x^{*}\right)=\lambda_{u}\left[1-p_{c}(x\mid x^{*})\right]$
and switching to polar coordinates for the integration. 

\section{Proof of Proposition \ref{prop:general bound}}

Replacing the PCA function in (\ref{eq:UE density}) with its lower
bound as per Cor. \ref{cor: PCA bound} results in the equivalent
density bound
\begin{align*}
 & \lambda_{x_{R},u}(r)\\
\leq & \lambda_{u}\left(1-\frac{1}{2\pi}\int_{0}^{2\pi}e^{-\lambda_{a}\pi|(r,\theta)+x_{R}-x^{*}|^{2}}d\theta\right)\\
\stackrel{(a)}{\leq} & \lambda_{u}\left(1-\frac{e^{-\lambda_{a}\pi|x^{*}|^{2}}}{2\pi}\int_{0}^{2\pi}e^{-\lambda_{a}\pi|(r,\theta)+x_{R}|^{2}}d\theta\right)\\
\stackrel{(b)}{=} & \lambda_{u}\left(1-\frac{e^{-\lambda_{a}\pi|x^{*}|^{2}}}{2\pi}\int_{0}^{2\pi}e^{-\lambda_{a}\pi|(r,\theta)+(|x_{R}|,0)|^{2}}d\theta\right)\\
\stackrel{(c)}{=} & \lambda_{u}\left(1-\frac{e^{-\lambda_{a}\pi(|x^{*}|^{2}+|x_{R}|^{2}+r^{2})}}{2\pi}\int_{0}^{2\pi}e^{\lambda_{a}\pi2r|x_{R}|\cos\theta}d\theta\right),
\end{align*}

\noindent where $(a)$ follows by application of the triangle inequality,
$(b)$ by noting that the integral is independent of $\angle x_{R}$
and $(c)$ since $|(r,\theta)+(|x_{R}|,0)|^{2}=r^{2}+|x_{R}|^{2}-2r|x_{R}|\cos(\pi-\theta)$
(cosine law). The integral in the last equation is equal to $2\pi\mathcal{I}_{0}(\lambda_{a}2\pi|x_{R}|r)$.
By exchanging the roles $x^{*}$ and $x_{R}$ in $(a)$ and following
the same reasoning, another upper bound of the form of $(c)$ results,
with the only difference that the integrand has $|x^{*}|$ instead
of $|x_{R}|$ and evaluates to $2\pi\mathcal{I}_{0}(\lambda_{a}2\pi|x^{*}|r)$.
Considering the minimum of these two bounds leads to (\ref{eq:general_upper_bound_for_density}).


\begin{thebibliography}{10}
\bibitem{Haenggi=000026Ganti}M. Haenggi and R. K. Ganti, \textquotedblleft Interference
in Large Wireless Networks,\textquotedblright{} \emph{Found. Trends
Netw}., vol. 3, no. 2, pp. 127\textendash 248, 2008.

\bibitem{Boccardi 5 disruptive}F. Boccardi, R. W. Heath, A. Lozano,
T. L. Marzetta, and P. Popovski, \textquotedblleft Five disruptive
technology directions for 5G,\textquotedblright{} \emph{IEEE Commun.
Mag}., vol. 52, pp. 74\textendash 80, Feb. 2014.

\bibitem{D2D general}K. Doppler, M. Rinne, C. Wijting, C. Ribeiro,
and K. Hugl, \textquotedblleft Device-to-Device Communication as an
Underlay to LTE-Advanced Networks,\textquotedblright{} \emph{IEEE
Commun. Mag}., vol. 50, no. 3, pp. 170\textendash 177, Mar. 2012.

\bibitem{FD JSAC tutorial}A. Sabharwal, P. Schniter, D. Guo, D. Bliss,
S. Rangarajan, and R. Wich- man, \textquotedblleft In-band full-duplex
wireless: Challenges and opportunities,\textquotedblright{}\emph{
IEEE J. Select. Areas Commun.}, vol. 32, pp. 1637\textendash 1652,
Sept. 2014.

\bibitem{D2D mag network controlled}L. Lei and Z. Zhong, \textquotedblleft Operator
controlled device-to-device communications in LTE-Advanced networks,\textquotedblright{}
\emph{IEEE Wireless Commun}., vol. 19, no. 3, pp. 96\textendash 104,
Jun. 2012.

\bibitem{JSAC stoc geom}M. Haenggi, J. G. Andrews, F. Baccelli, O.
Dousse, and M. Franceschetti, \textquotedblleft Stochastic geometry
and random graphs for the analysis and design of wireless networks,\textquotedblright{}
\emph{IEEE J. Select. Areas Commun.}, vol. 27, pp. 1029-1046, Sept.
2009.

\bibitem{D2D Ye}Q. Ye, M. Al-Shalash, C. Caramanis, and J. G. Andrews,
\textquotedblleft Resource optimization in device-to-device cellular
systems using time-frequency hopping,\textquotedblright{} \emph{IEEE
Trans. Wireless Commun}., vol. 13, no. 10, pp. 5467\textendash 5480,
Oct. 2014.

\bibitem{D2D Lin}X. Lin, J. G. Andrews, and A. Ghosh, \textquotedblleft Spectrum
sharing for device-to-device communication in cellular networks,\textquotedblright{}
\emph{IEEE Trans. Wireless Commun}., vol. 13, no. 12, pp. 6727\textendash 6740,
Dec. 2014. 

\bibitem{D2D Lee}N. Lee, X. Lin, J. G. Andrews, and R. W. Heath,
Jr., \textquotedblleft Power control for D2D underlaid cellular networks:
Modeling, algorithms and analysis,\textquotedblright{} \emph{IEEE
J. Sel. Areas Commun}., vol. 33, no. 1, pp. 1\textendash 13, Jan.
2015. 

\bibitem{Stefanatos}S. Stefanatos, A. Gotsis, and A. Alexiou, \textquotedblleft Operational
region of D2D communications for enhancing cellular network performance,\textquotedblright{}
\emph{IEEE Trans. Wireless Commun.}, vol. 14, no. 11, pp. 5984\textendash 5997,
Nov. 2015.

\bibitem{Andrews guard zone}A. Hasan and J. G. Andrews, \textquotedblleft The
guard zone in wireless ad hoc networks,\textquotedblright{} \emph{IEEE
Trans. Wireless Commun.,} vol. 6, no.3, pp. 897\textendash 906, Mar.
2007.

\bibitem{George D2D}G. George, R. K. Mungara, and A. Lozano, \textquotedblleft An
Analytical Framework for Device-to-Device Communication in Cellular
Networks,\textquotedblright{} \emph{IEEE Trans. Wireless Commun}.,
vol. 14, no. 11, pp. 6297\textendash 6310, Nov. 2015.

\bibitem{Kountouris D2D}Z. Chen and M. Kountouris, ``Decentralized
opportunistic access for D2D underlaid cellular networks,'' Jul.
2016. {[}Online{]}. Available: http://arxiv.org/abs/1607.05543.

\bibitem{Primer on PPP}J. G. Andrews, R. K. Ganti, N. Jindal, M.
Haenggi, and S. Weber, \textquotedblleft A primer on spatial modeling
and analysis in wireless networks,\textquotedblright{} \emph{IEEE
Commun. Mag.}, vol. 48, no. 11, pp. 156\textendash 163, Nov. 2010.

\bibitem{ElSawy uplink}H. ElSawy and E. Hossain, \textquotedblleft On
stochastic geometry modeling of cellular uplink transmission with
truncated channel inversion power control,\textquotedblright{}\emph{
IEEE Trans. Wireless Commun}., vol. 13, no. 8, pp. 4454\textendash 4469,
Aug. 2014.

\bibitem{UL Comm Lett}H. Lee, Y. Sang, and K. Kim, \textquotedblleft On
the uplink SIR distributions in heterogeneous cellular networks,\textquotedblright{}
\emph{IEEE Commun. Lett.}, vol. 18, no. 12, pp. 2145\textendash 2148,
Dec. 2014.

\bibitem{Andrews joint UL-DL}S. Singh, X. Zhang, and J. G. Andrews,
``Joint rate and SINR coverage analysis for decoupled uplink-downlink
biased cell associations in HetNets,'' \emph{IEEE Trans. Wireless
Commun.}, vol. 14, no. 10, pp. 5360\textendash 5373, Oct. 2015.

\bibitem{PVLP}B. B\l aszczyszyn and D. Yogeshwaran, \textquotedblleft Clustering
comparison of point processes with applications to random geometric
models,\textquotedblright{} in \emph{Stochastic Geometry, Spatial
Statistics and Random Fields}, vol. 2120, Lecture Notes in Mathematics.
Cham, Switzerland: Springer-Verlag, 2015, pp. 31\textendash 71.

\bibitem{Haenggi PVLP}M. Haenggi, ``User point processes in cellular
networks,'' \emph{IEEE Wireless Commun. Lett.}, vol. 6, no. 2, pp.
258\textendash 261, Apr. 2017.

\bibitem{DiRenzo}M. Di Renzo, W. Lu, and P. Guan, ``The intensity
matching approach: A tractable stochastic geometry approximation to
system-level analysis of cellular networks,'' \emph{IEEE Trans. Wireless
Commun.}, vol. 15, no. 9, pp. 5963\textendash 5983, Sep. 2016.

\bibitem{cross-cell D2D}H. Zhang, Y. Ji, L. Song, and H. Zhu \textquotedblleft Hypergraph
based resource allocation for cross-cell device-to-device communications,\textquotedblright{}
in \emph{Proc. IEEE Int. Conf. Commun. (ICC)}, Kuala Lumpur, Malaysia,
May 2016, pp. 1\textendash 6.

\bibitem{Tong=000026Haengi FD}Z. Tong and M. Haenggi, \textquotedblleft Throughput
analysis for full-duplex wireless networks with imperfect self-interference
cancellation,\textquotedblright{} \emph{IEEE Trans. Commun}., vol.
63, no. 11, pp. 4490\textendash 4500, Nov. 2015.

\bibitem{Psomas}C. Psomas, M. Mohammadi, I. Krikidis, and H. A. Suraweera,
``Impact of directionality on interference mitigation in full-duplex
cellular networks,'' \emph{IEEE Trans. Wireless Commun.}, vol. 16,
no. 1, pp. 487\textendash 502, Jan. 2017.

\bibitem{ElSawy FD with D2D}K. S. Ali, H. ElSawy, and M.-S. Alouini,
``Modeling cellular networks with full-duplex D2D communication:
a stochastic geometry approach,'' \emph{IEEE Trans. Commun}., vol.
64, no. 10, pp. 4409\textendash 4424, Oct. 2016.

\bibitem{Andrews tractable}J. G. Andrews, F. Baccelli, and R. K.
Ganti, \textquotedblleft A tractable approach to coverage and rate
in cellular networks,\textquotedblright{}\emph{ IEEE Trans. Commun}.,
vol. 59, no. 11, pp. 3122\textendash 3134, Nov. 2011.

\bibitem{Stefanatos UDN}S. Stefanatos and A. Alexiou, \textquotedblleft Access
Point density and bandwidth partitioning in ultra-dense wireless networks,\textquotedblright{}
\emph{IEEE Trans. Commun}., vol. 62, no. 9, pp. 3376\textendash 3384,
Sept. 2014.

\bibitem{Baccelli}F. Baccelli and B. B\l aszczyszyn, \emph{Stochastic
Geometry and Wireless Networks}. Delft, The Netherlands: Now Publishers
Inc., 2009.

\bibitem{Kountouris 2}Z. Chen and M. Kountouris, ``Guard zone based
D2D underlaid cellular networks with two-tier dependence,'' \emph{2015
IEEE International Conference on Communication Workshop (ICCW)}, London,
2015, pp. 222-227.

\bibitem{Haenggi local delay}M. Haenggi, \textquotedblleft The local
delay in poisson networks,\textquotedblright{} \emph{IEEE Trans. Inf.
Theory}, vol. 59, pp. 1788\textendash 1802, Mar. 2013. 

\bibitem{Dhillon hole}Z. Yazdanshenasan, H. S. Dhillon, M. Afshang,
and P. H. J. Chong, \textquotedblleft Poisson hole process: theory
and applications to wireless networks,\textquotedblright{} \emph{IEEE
Trans. Wireless Commun.}, vol. 15, no. 11, pp. 7531\textendash 7546,
Nov. 2016.

\bibitem{Handoff}S. Sadr and R. S. Adve, \textquotedblleft Handoff
rate and coverage analysis in multi-tier heterogeneous networks,\textquotedblright{}
\emph{IEEE Trans. Wireless Commun.}, vol. 14, no. 5, pp. 2626\textendash 2638,
May 2015.

\bibitem{Elkotby relay}H. E. Elkotby and M. Vu, \textquotedblleft Uplink
User-Assisted Relaying Deployment in Cellular Networks,\textquotedblright{}
\emph{IEEE Trans. on Wireless Commun.}, vol. 14, no. 10, pp. 5468\textendash 5483,
Oct. 2015.

\bibitem{Gamal}A. E. Gamal and Y.-H. Kim, \emph{Network Information
Theory}, Cambridge University Press, 2012.
\end{thebibliography}
\end{document}